\documentclass[twocolumn,preprintnumbers,amsmath,amssymb,superscriptaddress,nofootinbib,english]{revtex4-1}
\pdfoutput=1
\usepackage{times,amsmath,amsfonts,amssymb,epstopdf}
\usepackage{graphicx}
\usepackage{dcolumn}
\usepackage{bm}
\usepackage{enumerate}
\usepackage{epsfig}
\usepackage{graphicx}
\usepackage{hyperref}
\usepackage[usenames]{color}
\usepackage{url}
\usepackage[normalem]{ulem}
\usepackage[T1]{fontenc}
\usepackage{amsmath}

\usepackage[usenames]{color}

\def\be{\begin{equation}}
\def\ee{\end{equation}}
\def\ben{\begin{eqnarray}}
\def\een{\end{eqnarray}}
\def\ba{\begin{array}}
\def\ea{\end{array}}

\newcommand{\bq}{\begin{eqnarray}}
\newcommand{\eq}{\end{eqnarray}}
\newcommand{\bes}{\begin{subequations}}
\newcommand{\ees}{\end{subequations}}

\begin{document}
\newcommand{\half}{{\textstyle\frac{1}{2}}}
\allowdisplaybreaks[3]
\def\triangledown{\nabla}
\def\grad3{\hat{\nabla}}
\def\a{\alpha}
\def\b{\beta}
\def\g{\gamma}\def\G{\Gamma}
\def\d{\delta}\def\D{\Delta}
\def\ep{\epsilon}
\def\et{\eta}
\def\z{\zeta}
\def\t{\theta}\def\T{\Theta}
\def\l{\lambda}\def\L{\Lambda}
\def\m{\mu}
\def\f{\phi}\def\F{\Phi}
\def\n{\nu}
\def\r{\rho}
\def\s{\sigma}\def\S{\Sigma}
\def\ta{\tau}
\def\x{\chi}
\def\o{\omega}\def\O{\Omega}
\def\k{\kappa}
\def\pa {\partial}
\def\ov{\over}
\def\br{\\}
\def\ud{\underline}

\def\lcdm{\Lambda{\rm CDM}}
\def\qcdm{{\rm QCDM}}
\def\nloc{R\Box^{-2}R}
\def\msun{M_{\odot}/h}
\def\dw{f(X)}

\newcommand\lsim{\mathrel{\rlap{\lower4pt\hbox{\hskip1pt$\sim$}}
    \raise1pt\hbox{$<$}}}
\newcommand\gsim{\mathrel{\rlap{\lower4pt\hbox{\hskip1pt$\sim$}}
    \raise1pt\hbox{$>$}}}
\newcommand\esim{\mathrel{\rlap{\raise2pt\hbox{\hskip0pt$\sim$}}
    \lower1pt\hbox{$-$}}}
\newcommand{\dpar}[2]{\frac{\partial #1}{\partial #2}}
\newcommand{\sdp}[2]{\frac{\partial ^2 #1}{\partial #2 ^2}}
\newcommand{\dtot}[2]{\frac{d #1}{d #2}}
\newcommand{\sdt}[2]{\frac{d ^2 #1}{d #2 ^2}}    

\title{Weak lensing by voids in modified lensing potentials}

\author{Alexandre Barreira}
\email[Electronic address: ]{a.m.r.barreira@durham.ac.uk}
\affiliation{Institute for Computational Cosmology, Department of Physics, Durham University, Durham DH1 3LE, U.K.}
\affiliation{Institute for Particle Physics Phenomenology, Department of Physics, Durham University, Durham DH1 3LE, U.K.}

\author{Marius Cautun}
\affiliation{Institute for Computational Cosmology, Department of Physics, Durham University, Durham DH1 3LE, U.K.}

\author{Baojiu Li}
\affiliation{Institute for Computational Cosmology, Department of Physics, Durham University, Durham DH1 3LE, U.K.}

\author{Carlton M. Baugh}
\affiliation{Institute for Computational Cosmology, Department of Physics, Durham University, Durham DH1 3LE, U.K.}

\author{Silvia Pascoli}
\affiliation{Institute for Particle Physics Phenomenology, Department of Physics, Durham University, Durham DH1 3LE, U.K.}

\preprint{IPPP/15/ 31 DCPT/15/ 62}

\begin{abstract}
We study lensing by voids in Cubic Galileon and Nonlocal gravity cosmologies, which are examples of theories of gravity that modify the lensing potential. We find voids in the dark matter and halo density fields of N-body simulations and compute their lensing signal analytically from the void density profiles, which we show are well fit by a simple analytical formula. In the Cubic Galileon model, the modifications to gravity inside voids are not screened and they approximately double the size of the lensing effects compared to GR. The difference is largely determined by the direct effects of the fifth force on lensing and less so by the modified density profiles. For this model, we also discuss the subtle impact on the force and lensing calculations caused by the screening effects of haloes that exist in and around voids. In the Nonlocal model, the impact of the modified density profiles and the direct modifications to lensing are comparable, but they boost the lensing signal by only $\approx 10\%$, compared with that of GR. Overall, our results suggest that lensing by voids is a promising tool to test models of gravity that modify lensing.

\end{abstract} 
\maketitle
\section{Introduction}\label{sec:into}

Despite the success of General Relativy (GR) in passing all currently available solar system tests of gravity \cite{Will:2014xja},  there is growing interest in the theoretical \cite{Joyce:2014kja, 2012PhR...513....1C} and observational \cite{Jain:2007yk, Jain:2013wgs, Koyama:2015vza} aspects of theories beyond GR. There are two main reasons for this. Firstly, the simple fact that GR has not been tested on scales larger than the solar system means that, in fact, one makes a huge extrapolation of the regime of validity of the theory when one uses it (as it is common) in cosmological studies. The gravitational law should, therefore, be put to test on larger scales, and {\it modified gravity} models help to identify the types of imprints that modifications to gravity can leave on observables. Secondly, there is currently no theoretically appealing explanation for the nature of the {\it dark energy} that is responsible for the accelerated expansion of the Universe. In the standard $\Lambda$-Cold Dark Matter ($\Lambda$CDM) cosmological model, the role of the dark energy is attributed to a simple cosmological constant $\Lambda$, but the smallness of its value remains a mystery. Models of modified gravity can explain the acceleration without $\Lambda$, thereby providing extra motivation for their study.

The majority of modified gravity models predict the existence of extra degrees of freedom (often of the scalar type) that mediate {\it fifth forces} felt by the matter fields. Consequently, a major difficulty in building models of modified gravity comes from making them compatible with the stringent solar system bounds. The latter constrain the fifth force to be extremelly small, and hence, cosmologically uninteresting. A popular way out of this relies in building models that possess what are commonly referred to as {\it screening mechanisms}. In short, the idea is to construct models where the equations of the scalar field become highly nonlinear in regions of high density (like the solar system). The presence of the nonlinearities acts to suppress the magnitude of the fifth force. On larger scales, where the density is low, the fifth force effects become manifest and potentially detectable. On these large scales, the scalar field equation can be linearized to look like a Poisson equation. Examples of screening mechanisms include the {\it chameleon mechanism} \cite{Khoury:2003aq} which operates in the popular Hu-Sawicki $f(R)$ \cite{Hu:2007nk} gravity model; the {\it Vainshtein mechanism} \cite{Vainshtein1972393, Koyama:2013paa, Babichev:2013usa} which operates in the Dvali-Gabadadze-Porrati (DGP) \cite{Dvali:2000hr} and Galileon \cite{PhysRevD.79.064036, PhysRevD.79.084003, Deffayet:2009mn} models; the {\it K-mouflage screening} \cite{Babichev:2009ee, Brax:2014wla, Brax:2014yla, Brax:2014gra, Barreira:2014gwa} and {\it disformal screening} \cite{2012PhRvL.109x1102K, Sakstein:2014isa}.

Due to the suppression effects of the screening, it is best to devise observational tests that focus on large scales or low-density regions, where the screening is less efficient \cite{Falck:2014jwa, Falck:2015rsa}. For instance, recent studies have shown that the amplitude of the cosmic microwave background (CMB) lensing potential \cite{2012PhRvD..86l4016B, Barreira:2014jha, Barreira:2014gwa, Ade:2015rim} and cosmic shear \cite{Battye:2014xna, Leonard:2015hha} power spectra (which are sensitive to the projected matter distribution on large scales) are, indeed, a sensitive probe of modified gravity. The cross-correlation of galaxy positions with the lensing shear field can also help to constrain modified gravity \cite{2011PhRvL.106t1102W, Park:2014aga}. The integrated Sachs-Wolfe (ISW) effect, which probes the time variation of large scale gravitational potentials, constitutes another good example of constraining gravity away from the regimes where the screening is at play \cite{2012PhRvD..86l4016B, Barreira:2014jha, Munshi:2014tua, Soergel:2014sna, Cai:2013toa, 2012PhRvD..85l3503K, 2009PhRvD..80f3536L}. On the other hand, although the amplitude of the matter power spectrum on large scales is also affected by the modifications to gravity, the uncertainties about galaxy bias undermine the possibility of obtaining tight constraints (see e.g.~Sec.~IV.~D of Ref.~\cite{Barreira:2014zza} for a discussion). On mildly nonlinear scales ($2-20{\rm Mpc}$), several recent studies have found that the peculiar velocities of galaxies are also very sensitive to the presence of fifth forces \cite{Hellwing:2014nma, 2012PhRvL.109e1301L, Lam:2013kma, Zu:2013joa}. These scales are typically associated with the infall regions of massive galaxy clusters, which are located sufficiently far away from the cluster center for the screening to have a smaller impact. In general, inside the virial radius of galaxy clusters ($\lesssim 1{\rm Mpc}$), it becomes harder to find the effects of the fifth force (see e.g.~Ref.~\cite{Barreira:2015fpa}).

Here, we focus on cosmic voids, which are the regions of the Universe where the density is the lowest, and hence, where one expects fifth force effects to be maximal. Despite being potentially good probes of gravity, voids have only recently become the object of dedicated studies in modified gravity \cite{2011MNRAS.411.2615L,2012MNRAS.421.3481L, 2013MNRAS.431..749C, Lam:2014kua, Cai:2014fma, Zivick:2014uva}. In particular, Ref.~\cite{Cai:2014fma} showed that the lensing signal from voids in $f(R)$ gravity is modified relative to $\Lambda$CDM, via the modifications induced by the fifth force to the void density profiles. This result is particularly timely as it can be linked to the recent work of Refs.~\cite{Melchior:2013gxd, Clampitt:2014gpa}, who have independently detected the lensing signal associated with cosmic voids. This therefore opens the prospect of developing new tests of gravity using the lensing signal in and around voids. In terms of lensing, $f(R)$ models (and scalar-tensor theories in general) are special in the sense that they do not modify the lensing signal directly. In these models, the amplitude of the fifth force vanishes for relativistic particles like photons. In other words, any modifications to lensing arise through changes in the mass distribution, and not due to changes to the photon geodesic equation. For this reason, one expects that lensing by voids can serve as a stronger probe of models that also modify the photon geodesic equation. Examples of such models include Nonlocal gravity \cite{Deser:2007jk, Deser:2013uya, Maggiore:2014sia, Dirian:2014ara, Dirian:2014bma, Barreira:2014kra}, Galileon gravity \cite{PhysRevD.79.064036, PhysRevD.79.084003, Deffayet:2009mn}, massive gravity \cite{Gabadadze:2009ja, deRham:2009rm, deRham:2010kj, deRham:2010gu, deRham:2010ik, deRham:2010tw,  Hassan:2011zd}, K-mouflage gravity \cite{Brax:2014wla, Brax:2014yla, Barreira:2014gwa, Babichev:2009ee}, Kinetic Gravity Braiding \cite{Deffayet:2010qz, Pujolas:2011he, 2012PhRvD..85l3503K} and several other special cases of Horndeski's general model \cite{Horndeski:1974wa}. One of our goals here is to investigate the lensing signal from voids in some of these models.

As working cases, we focus on the Cubic Galileon model \cite{Barreira:2014ija} and the Nonlocal gravity model of Ref.~\cite{Maggiore:2014sia}. We make use of the N-body simulations performed for these two models in Refs.~\cite{Barreira:2013eea, Barreira:2014kra}. We find voids in the simulations using a {\it watershed} based algorithm \cite{Platen:2007qk} and investigate the effects of the fifth force on the number of voids and on their density and force profiles. We also put forward a simple fitting formula that matches very well the void profiles found in the simulations for different variants of the modified gravity models, for different density tracer types (dark matter and haloes) and for a wide range of void sizes. Our formula is a simple extension of others used previously \cite{Hamaus:2014fma, Nadathur:2014qja}, and by having more parameters it provides a better fit to our simulation results. The formula admits a closed expression (in terms of hypergeometric functions) for the mass within a given radius, which makes it convenient to use in force profile calculations and lensing studies.  When we assess the impact of the fifth force on the lensing signal, we take into account its effect on both the void density profiles and the calculation of the lensing observables themselves. Our goal is to provide intuition about the potential of lensing by voids to test gravity outside the solar system. We do not attempt to make any observationally conclusive statement, but we do comment on a number of extra steps that need to be taken to compare our results with observations.

This paper is organized as follows. In Sec.~\ref{sec:themodels}, we introduce the force equations in Nonlocal and Galileon gravity, discussing some of their phenomenology. In Sec.~\ref{sec:voidsims}, we describe our N-body simulations and the void finding algorithm, and study the effects of the fifth force on the abundance, density profiles and force profiles of the voids. In Sec.~\ref{sec:lensing}, we describe the calculation of the lensing signal, and then compute it for the voids found in the simulations. In the same section, we also link our findings to recent observational results, and provide a quick guideline of the steps needed for more elaborate comparisons to observations. We summarize our results in Sec.~\ref{sec:summary}.


\section{The models of gravity}\label{sec:themodels}

In this section, we briefly introduce the models of gravity that we consider and present the relevant force equations that are needed to compute their lensing signal. In the equations below we always assume spherical symmetry and work with a perturbed Friedmann-Robertson-Walker (FRW) spacetime in the Newtonian gauge
\bq\label{eq:metric}
{\rm d}s^2 = \left(1 + 2\Psi/c^2\right)c^2{\rm d}t^2 - a^2\left(1 - 2\Phi/c^2\right){\rm d}{\bf x}^2,
\eq
where $a = 1/(1+z)$ is the cosmological scale factor ($z$ is the redshift) and $c$ is the speed of light.

\subsection{Nonlocal gravity}\label{sec:nloc}

We consider the Nonlocal gravity model of Refs.~\cite{Maggiore:2014sia, Dirian:2014ara}. Its action is given by

\bq\label{eq:action}
S = \frac{1}{16 \pi G}\int {\rm d}x^4\sqrt{-g}\left[\mathcal{R} - \frac{m^2}{6}\mathcal{R}\Box^{-2}\mathcal{R} - \mathcal{L}_m\right],
\eq
which can be cast in a more familiar (local) form given by \cite{Nojiri:2007uq, Capozziello:2008gu, Koshelev:2008ie}

\bq\label{eq:action-local}
A &&= \frac{1}{16 \pi G}\int {\rm d}x^4\sqrt{-g}\left[\mathcal{R} - \frac{m^2}{6}\mathcal{R}S - \xi_1\left(\Box U + \mathcal{R}\right) \right. \nonumber \\
&&\ \ \ \ \ \ \ \ \ \ \ \ \  \ \ \ \ \ \ \ \ \ \ \ \ \ \ \ \ \ \ \ \left. - \xi_2\left(\Box S + U\right) - \mathcal{L}_m\right],
\eq
where $\mathcal{R}$ is the Ricci scalar, $g$ is the determinant of the metric $g_{\mu\nu}$, $G$ is Newton's gravitational constant, $\mathcal{L}_m$ is the matter Lagrangian density, $\xi_1$ and $\xi_2$ are Lagrange multipliers, $U = - \Box^{-1} \mathcal{R}$ and $S = \Box^{-2}\mathcal{R}$ are two auxiliary scalar fields and $\Box = \nabla^\mu\nabla_\mu$ is the d'Alembert operator, with Greek indices running over $0$, $1$, $2$, $3$. Here, we do not present a detailed discussion about the theoretical aspects of the above two actions, but simply caution that their solutions are not completely equivalent and that care must be taken before matching them (see, e.g.~Refs.~\cite{Koshelev:2008ie, Koivisto:2009jn, Barvinsky:2011rk, Deser:2013uya, Maggiore:2013mea, Foffa:2013sma, Foffa:2013vma} for a discussion).

On the scales relevant for large scale structure formation and in the absence of anisotropic stress, the two Newtonian potentials are the same ($\Psi = \Phi$) and the modifed Poisson equation can be written as \cite{Dirian:2014ara, Barreira:2014kra}
\bq\label{eq:modpoisson}
\frac{1}{R^2}\left(R^2\Phi,_R\right),_R = 4\pi G_{\rm eff} \bar{\rho}_m\delta(R),
\eq
where $\bar{\rho}_m$ is the cosmological background value of the physical matter density ${\rho}_m$, $\delta = \rho_m/\bar{\rho}_m - 1$ is the density contrast and $,_R$ denotes partial differentiation w.r.t.~the radial coordinate $R$. The above equation has the same form as in GR but with an effective time-dependent gravitational strength given by

\bq\label{eq:geff}
G_{\rm eff} = G \left[1 - \frac{m^2\bar{S}(z)}{3}\right]^{-1} \geq 1,
\eq
where $\bar{S}$ is the background part of the field $S$. The time evolution of the background quantities in the Nonlocal model have to be obtained numerically by integrating the background differential equations (see e.g.~Refs~\cite{Dirian:2014ara, Barreira:2014kra}). The parameter $m$ in Eqs.~(\ref{eq:action}) and (\ref{eq:action-local}) is controlled by the amount of dark energy in the Universe, i.e., in a flat Universe, its value is determined by the energy densities of the remaining matter species. This means that this Nonlocal gravity model has the same number of free parameters as $\lcdm$. For reference, for the model parameters used in Ref.~\cite{Barreira:2014kra} (e.g.~$\Omega_{m0} = 0.30$), one has that $G_{\rm eff}(z=0)/G \approx 1.06$.

The fact that this model is characterized by an enhanced gravitational strength on all length scales leads to the question of whether or not this model is capable of passing solar system constraints \cite{Will:2014xja}. In Ref.~\cite{Barreira:2014kra}, we showed that if the gravitational strength of Eq.~(\ref{eq:geff}) is used in solar system tests, then the model predicts values for the rate of change of the gravitational strength,  $\dot{G}_{\rm eff}$, that are incompatible with current lunar laser ranging experiments \cite{Williams:2004qba}. However, the time evolution of $G_{\rm eff}/G$ is controled by the background part of the field $S$. This means that if one describes the spacetime around the Sun as perturbed Minkowskii (instead of FRW), then $\dot{\bar{S}} = 0 \rightarrow \dot{G}_{\rm eff} = 0$, rendering the model compatible with current bounds \cite{Kehagias:2014sda, Maggiore:2014sia}. Here, we focus on void size scales, which are sensitive to the background expansion, and as a result, we use the gravitational strength of Eq.~(\ref{eq:geff}) when computing the model predictions.

\subsection{Cubic Galileon gravity}\label{sec:gali}

\begin{figure}
	\centering
	\includegraphics[scale=0.345]{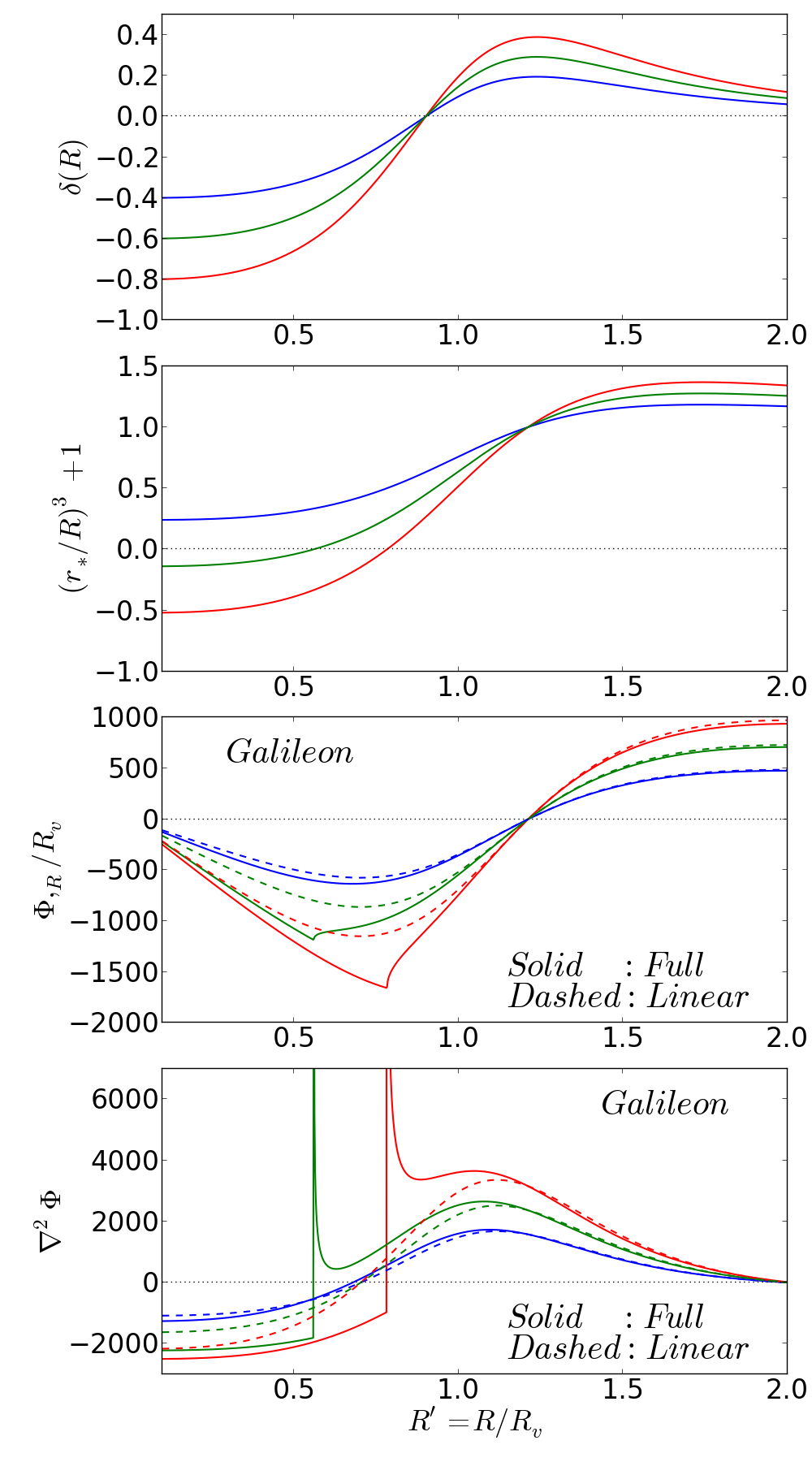}
	\caption{Representative force profiles in the Cubic Galileon model. The top panel shows three example void density profiles. The three lowest panels, from top to bottom,  show the radial profiles of $(r_*/R)^3 + 1$, $\Phi,_R$ and $\nabla^2\Phi$, for the density profiles shown in the top panel. The colors indicate which prediction is associated with which density profile. In the bottom two panels, the dashed and solid lines correspond, respectively, to the predictions of the linearized (cf.~Eq.~(\ref{eq:linforcegali}), regime (ii)) and full (cf.~Eqs.~(\ref{eq:phiror}) and (\ref{eq:phirr})) Cubic Galileon models, as labelled. The meaning of a negative amplitude for the force is that it points outwards. The units in the bottom two panels are $({\rm km/s})^2 h^2 {\rm Mpc}^{-2}$.}
\label{fig:complex} \end{figure}

We focus on the Cubic sector of the covariant Galileon gravity model \cite{PhysRevD.79.064036, PhysRevD.79.084003, Deffayet:2009mn, 2010PhRvL.105k1301D} whose action is given by
\bq\label{Galileon action}
&& S = \int {\rm d}^4x\sqrt{-g} \left[\frac{\mathcal{R}}{16\pi G} - \frac{1}{2}c_2\mathcal{L}_2 - \frac{1}{2}c_3\mathcal{L}_3 - \mathcal{L}_m\right], \nonumber \\
\eq
where $c_2$ and $c_3$ are dimensionless constants, and $\mathcal{L}_2$ and $\mathcal{L}_3$ are given by
\bq\label{L's}
\mathcal{L}_2 = \nabla_\mu\varphi\nabla^\mu\varphi,\ \ \ \ \ \ \ \  \mathcal{L}_3 = \frac{2}{\mathcal{M}^3}\Box\varphi\nabla_\mu\varphi\nabla^\mu\varphi,
\eq
in which $\varphi$ is the Galileon field, $\mathcal{M}^3 = M_{\rm Pl}H_0^2$, $M_{\rm Pl}^2 = 1/(8 \pi G)$ is the reduced Planck mass squared and $H_0 = 100 h\ {\rm km/s/Mpc}$ is the present-day Hubble expansion rate. In flat spacetime, the above action is invariant under the {\it Galilean shift} $\partial\varphi \rightarrow \partial\varphi + b_\mu$ (where $b_\mu$ is a constant four-vector). Following the derivation of Refs.~\cite{Barreira:2013eea, Barreira:2013xea}, the force law in the Cubic Galileon model is given by 

\bq\label{eq:intpoissongali}
\frac{\Phi,_R}{R} = \frac{G\delta M(<R)}{R^3} - \frac{c_3}{\mathcal{M}^3}\dot{\bar{\varphi}}^2\frac{\delta\varphi,_R}{R},
\eq
where $\delta\varphi$ is the spatial perturbation of the Galileon field, $\bar{\varphi}(z)$ is its backround part, and $\delta M(<R) = 4\pi \bar{\rho}_m\int_0^R \delta(x)x^2{\rm d}x$ is the mass perturbation enclosed in a sphere of radius $R$. Compared to GR, Eq.~(\ref{eq:intpoissongali}) has an extra term, which is proportional to 

\bq\label{eq:galieomsol}
\frac{\delta\varphi,_R}{R} = \frac{4}{3}\frac{M_{\rm Pl}}{\beta_2}\left(\frac{R}{r_*}\right)^3\left[\sqrt{\left(\frac{r_*}{R}\right)^3 + 1} - 1\right]\frac{G\delta M(<R)}{R^3}, \nonumber \\
\eq
with
\bq\label{eq:rv}
r_*^3 = \frac{16}{9}\frac{M_{\rm Pl}}{\beta_1\beta_2\mathcal{M}^3}G\delta M(<R),
\eq
where $\beta_1$ and $\beta_2$ are two dimensionless functions of time. The quantity $r_*$ is a radial scale, which is often referred to as the Vainshtein radius. From Eqs.~(\ref{eq:intpoissongali}), (\ref{eq:galieomsol}), one can write

\begin{widetext}
\bq
\label{eq:phiror}\frac{\Phi,_R}{R} &=& \left\{1 - \frac{4}{3}\frac{c_3}{M_{\rm Pl}\mathcal{M}^3}\frac{\dot{\bar{\varphi}}^2}{\beta_2}\left(\frac{R}{r_*}\right)^3\left[\sqrt{\left(\frac{r_*}{R}\right)^3 + 1} - 1\right]\right\}\frac{G\delta M(<R)}{R^3}, \\
\nonumber \\
\label{eq:phirr}\Phi,_{RR} &=& G\left[\frac{\delta M(<R),_R}{R^2} - \frac{2\delta M(<R)}{R^3}\right] - \frac{3}{4}\frac{c_3\beta_1\dot{\bar{\varphi}}^2}{M_{\rm Pl}^2}\left[\sqrt{\left(\frac{r_*}{R}\right)^3 + 1} - 1 + \frac{3}{2}\frac{(r_*/R)^2}{\sqrt{(r_*/R)^3 + 1}}\left(r_*,_R - \frac{r_*}{R}\right)\right].
\eq
\end{widetext}

Contrary to the case of Nonlocal gravity, the Galileon model admits analytical solutions for the time evolution of the background quantities \cite{Barreira:2014jha}. The time evolution of the Hubble parameter, $\dot{\bar{\varphi}}$, $\beta_1$ and $\beta_2$ are given, respectively, by 

\bq\label{eq:galibg}
H^2 &=& \frac{H_0^2}{2}\left[\Omega_{m0}a^{-3} + \sqrt{\Omega_{m0}^2a^{-6}  + 4(1 - \Omega_{m0})}\right], \\
\dot{\bar{\varphi}} &=& \xi H_0^2/H, \\
\beta_1 &=& \frac{1}{6c_3}\left[-c_2 - \frac{4c_3}{\mathcal{M}^3}\left(\ddot{\bar{\varphi}} + 2H\dot{\bar{\varphi}}\right) + \frac{2 c_3^2}{M_{\rm Pl}^2\mathcal{M}^6}\dot{\bar{\varphi}}^4\right], \\
\beta_2 &=& \frac{2\mathcal{M}^3M_{\rm Pl}}{\dot{\bar{\varphi}}^2}\beta_1.
\eq
As in Ref.~\cite{Barreira:2014jha}, we take $c_2 = -1$ and the other two Galileon parameters are determined by $\Omega_{m0}$ as

\bq\label{eq:c3ksi}
\xi &=& \sqrt{6(1-\Omega_{m0})}, \\
c_3 &=&  1/(6\xi).
\eq
We take $\Omega_{m0} = 0.28$, which is the value used in the simulations of Ref.~\cite{Barreira:2013eea}.

From Eq.~(\ref{eq:phiror}), it is possible to identify three regimes for the amplitude of the total force in the Cubic Galileon model that are relevant for our analysis:

\setlength{\leftmargini}{10pt}\begin{enumerate}[i]

\item In the regime where $r_* > 0 $ and ${r_*}/{R} \gg 1$, one has

\bq
\left(\frac{R}{r_*}\right)^3\left[\sqrt{\left(\frac{r_*}{R}\right)^3 + 1} - 1\right] \approx \left(\frac{r_*}{R}\right)^{-3/2} \ll 1,
\eq
and, as a result, Eq.~(\ref{eq:phiror}) can be approximated as 

\bq
\frac{\Phi,_R}{R} \approx \frac{G\delta M(<R)}{R^3}.
\eq
That is, close to very massive objects (small $R$ and/or large mass perturbations, $r_* \propto \delta M$), the force law in the Galileon model becomes the same as in GR. This illustrates the implementation of the Vainshtein screening effect that allows this model to satisfy solar system tests of gravity.

\item If $r_* > 0$ and ${r_*}/{R} \ll 1$, then 
\bq\label{eq:linapprox}
\left(\frac{R}{r_*}\right)^3\left[\sqrt{\left(\frac{r_*}{R}\right)^3 + 1} - 1\right] \approx 1/2,
\eq
and Eq.~(\ref{eq:phiror}) becomes
\bq\label{eq:linforcegali}
\frac{\Phi,_R}{R} &=& G_{\rm eff}(z)\frac{\delta M(<R)}{R^3}, \\ 
\ \nonumber \\
G_{\rm eff} &=& G\left(1 - \frac{2}{3}\frac{c_3}{M_{\rm Pl}\mathcal{M}^3}\frac{\dot{\bar{\varphi}}^2}{\beta_2}\right) > 1
\eq
($\beta_2 < 0$ \cite{Barreira:2013eea}). In this {\it linear} regime, which occurs sufficiently far away from massive objects, the force law is as in GR, but with an enhanced time-dependent gravitational strength. This is similar to the force law of the Nonlocal gravity model, albeit with a different time evolution for $G_{\rm eff}$. In particular, in the Galileon model, and for the model parameters used in Ref.~\cite{Barreira:2013eea}, one has $G_{\rm eff}(z = 0)/G \approx 2$, which is subtantially stronger than the $\approx 6\%$ enhancement in the Nonlocal model.

\item Finally, there is a third regime characterized by $r_* < 0$ and $|r_*/R| \sim \mathcal{O}(1)$. In this regime, which occurs whenever the mass perturbation becomes negative (as it does in voids), the total force can be written as
\bq
\frac{\Phi,_R}{R} &=& G_{\rm void}(z, R)\frac{\delta M(<R)}{R^3},
\eq
where $G_{\rm void}(z, R)$ is a time and scale dependent effective gravitational strength (simply the term between $\left\{\right\}$ in Eq.~(\ref{eq:phiror})), which is larger in magnitude than the gravitational strength of regime (ii), i.e., $G_{\rm void} > G_{\rm eff}$. This can be checked by noting that 
\bq
\left(\frac{R}{r_*}\right)^3\left[\sqrt{\left(\frac{r_*}{R}\right)^3 + 1} - 1\right] > 1/2
\eq
 in Eq.~(\ref{eq:phiror}), when $r_* < 0$ (cf.~Eq.~(\ref{eq:linapprox})). Note that in our notation, when $\delta M < 0$, then the force becomes negative. This means that the force points outwards.

\end{enumerate}

To help understand the behavior of the fifth force in the Galileon model, we show in Fig.~\ref{fig:complex}, the radial profiles of $(r_*/R)^3 +1$, $\Phi,_R$ and $\nabla^2\Phi = \Phi,_{RR} + 2\Phi,_R/R$, for each of the density profiles depicted in the top panel. The density profiles are computed using the formula

\bq\label{eq:mine}
\delta(R' = R/R_v) = \delta_v \frac{1 - \left(R'/s_1\right)^\alpha}{1 + \left(R'/s_2\right)^\beta},
\eq
where $R_v$ is the void radius (whose exact value is not important for the discussion here) and $\delta_v$, $\alpha$, $\beta$, $s_1$ and $s_2$ are fitting parameters. Figure \ref{fig:voidparamsimpact} in the Appendix shows the impact that each of the five parameters of Eq.~(\ref{eq:mine}) has on the density profiles (and on the associated lensing signal, whose calculation is explained in Sec.~\ref{sec:calculation}). In Sec.~\ref{sec:profiles}, we shall see that this formula provides a very good fit to the void density profiles found in the N-body simulations. The mass perturbation, $\delta M(<R) = 4\pi\bar{\rho}_m\int_0^R \delta(x)x^2{\rm d}x$, admits a closed formula given by

\begin{widetext}
\bq\label{eq:mass}
\delta M(<R) = 4 \pi \bar{\rho}_m\frac{R^3}{3(\alpha+3)} \delta_v \left[(\alpha + 3)\ _2F_1\left(1, \frac{3}{\beta}, \frac{\beta+3}{\beta}, -\left(\frac{R'}{s_2}\right)^\beta\right) - 3\left(\frac{R'}{s_1}\right)^\alpha\ _2F_1\left(1, \frac{\alpha+3}{\beta}, \frac{\alpha+\beta+3}{\beta}, -\left(\frac{R'}{s_2}\right)^\beta\right)\right], \nonumber \\
\eq
\end{widetext}
where $_2F_1$ is the Gauss hypergeometric series function. This formula for $\delta M(<R)$ facilitates straightforward calculation of the force profiles. In the bottom two panels of Fig.~\ref{fig:complex}, the solid curves show the result obtained by using Eqs.~(\ref{eq:phiror}) and (\ref{eq:phirr}) (which we call the {\it full} solution), whereas the dashed curves show the result associated with the regime (ii) above (Eq.~(\ref{eq:linforcegali}), which we call the {\it linear} solution). For the cases shown, for $R' \gtrsim 1.2$ ($r_* > 0$), the full and the linear solutions for the total force, $\Phi,_R$, are roughly the same, which is as expected since these radial scales correspond to the linear regime (ii) discussed above. A more careful inspection of those scales shows that the solid curves underpredict slightly the dashed ones. This is due to the Vainshtein screening mechanism, which acts to somewhat suppress the full solution. On the other hand, for $R' \lesssim 1.2$ ($r_* < 0$), the full solution enters regime (iii), and as expected, the force becomes larger in magnitude (more negative) compared to the linear case. 

There is one peculiar aspect about regime (iii) that is worth discussing with more detail. The quantity $(r_*/R)^3 +1$ appears in Eqs.~(\ref{eq:phiror}) and (\ref{eq:phirr}) as the argument of square-roots. This implies that the amplitude of the force becomes a complex number whenever $(r_*/R)^3 +1 < 0$, which is not a physical result. This problem has been already encountered in the N-body simulations of Ref.~\cite{Barreira:2013eea}, where the authors circunvent the absence of real solutions by adopting the {\it ad-hoc} fix of setting $(r_*/R)^3 +1 = 0$ whenever it becomes negative. Here, we shall implement the same procedure, as we wish to compare some of our results to those of Ref.~\cite{Barreira:2013eea}. In Fig.~\ref{fig:complex}, the implementation of this "fix" is noted by the {\it kinks} in $\Phi,_R$ and {\it spikes} in $\nabla^2\Phi$, for the solid green and solid red curves. The density inside the void depicted by the blue line is not low enough for $(r_*/R)^3 +1$ to cross zero, and hence, the problem is not seen.

It is important to try to understand the implications of the existence of complex solutions for the fifth force in the Galileon model. Here, we note that Eqs.~(\ref{eq:phiror}) and (\ref{eq:phirr}) are obtained under the approximations that the perturbed fields are weak and quasi-static. In the weak field approximation, one neglects terms that involve the perturbed fields and their first spatial derivatives, over their second derivatives. The quasi-static approximation amounts to neglecting the time variation of the perturbed quantities\footnote{These approximations are also used in the Nonlocal gravity model to obtain Eq.~(\ref{eq:modpoisson}), but in this model the problem of complex solutions does not arise.}, e.g., $\dot{\varphi}(a,R) = \dot{\bar{\varphi}}(a) + \dot{\delta\varphi}(a, R) \approx \dot{\bar{\varphi}}(a)$. In the very low-density regions that characterize voids, one expects the weak-field approximation to still hold, but the quasi-static one may not (see Refs.~\cite{Noller:2013wca, Sawicki:2015zya} for discussions about the quasi-static limit, and Refs.~\cite{Llinares:2013jua, Bose:2014zba} for work beyond this in N-body simulations). One may speculate that the terms which are neglected in the quasi-static limit are actually responsible for keeping the fifth force real for all density values when they are present. Interestingly, however, the recent work of Ref.~\cite{Winther:2015pta} has shown that the problem remains even after relaxing the quasi-static approximation. In particular, the authors find that the time derivative of the Galileon field perturbation becomes singular when the quasi-static solution becomes a complex number (see Ref.~\cite{Winther:2015pta} for the details). This suggests that the breakdown of the quasi-static solutions may well be associated with a true instability of the Cubic Galileon model. Moreover, the earlier work of Ref.~\cite{2014PhRvD..90l4035B} also unveiled some instabilities in general Vainshtein solutions around static spherically symmetric sources (see Ref.~\cite{2014PhRvD..90l4035B} for the details). Here, we shall keep these discussions in mind but proceed by retaining the {\it ad hoc} fix of Ref.~\cite{Barreira:2013eea}. Our treatment of the equations of Cubic Galileon can be viewed as a toy model that we use to illustrate the effects of modified gravity in the properties of voids.

\section{Voids in the simulations}\label{sec:voidsims}

\subsection{Outline of the simulations and void finder}\label{sec:counts}

\begin{table}
\caption{Summary of the model variants of the Cubic Galileon and Nonlocal models studied in this paper.}
\begin{tabular}{@{}lccccccccccc}
\hline\hline
\\
Model  & \ \ Expansion history & Force law&\ \ 
\\
\hline
\\
Full Galileon                    &\ \  Galileon &  \ \ Eqs.~(\ref{eq:phiror}) and (\ref{eq:phirr}) &\ \ 
\\
Linear Galileon               &\ \ Galileon &  Eq.~(\ref{eq:linforcegali}) &\ \ 
\\
$\rm{QCDM}_{\rm Galileon}$   &\ \ Galileon &  GR&\ \ 
\\
\\
Full Nonlocal                    &\ \  Nonlocal &  Eq.~(\ref{eq:modpoisson})  &\ \ 
\\
$\rm{QCDM}_{\rm Nonlocal}$   &\ \ Nonlocal &  GR&\ \ 
\\
\hline
\hline
\end{tabular}
\label{table-variants}
\end{table}

We make use of the N-body simulations of the Cubic Galileon and Nonlocal gravity models presented in Refs.~\cite{Barreira:2013eea, Barreira:2014kra}. For the case of the Nonlocal model, the simulations were run with a modified version of the Adaptive Mesh Refinement (AMR) and publicly-available {\tt RAMSES} code \cite{Teyssier:2001cp}. The modifications involved (i) changing the code to interpolate the expansion rate in the Nonlocal model using a table; (ii) rescaling the gravitational force computed in the default code by the value of $G_{\rm eff}/G$ in Eq.~(\ref{eq:geff}), which is also interpolated from tabulated values. For the Cubic Galileon model, the simulations were performed with the {\tt ECOSMOG} code \cite{Li:2013nua}, which is also based on {\tt RAMSES}. The {\tt ECOSMOG} code contains additional subroutines that solve the equation of motion of the Galileon scalar field via Newton-Gauss-Seidel iterative relaxations on the AMR grid. Once the values of the Galileon field are found on the grid, its spatial gradient is obtained by finite differencing to determine the fifth force. The modified background expansion history is also consistently taken into account by the {\tt ECOSMOG} code. We refer the reader to Refs.~\cite{Li:2013nua, Li:2013tda} for more details about how to solve Galileon-like equations in adaptively refined grids (see also Refs.~\cite{2009PhRvD..80d3001S, 2009PhRvD..80l3003S}).

Hereonin, we shall analyse the results from three variants of the Cubic Galileon model and two variants of the Nonlocal model, as listed in Table \ref{table-variants}. For the Galileon, we call these the {\it full}, {\it linear} and {\it QCDM} variants. The {\it QCDM model} is characterized by having the background expansion of the Galileon model, but the gravitational law of GR. The {\it linear model} is the same as QCDM, but with the effective gravitational strength $G_{\rm eff}$ of Eq.~(\ref{eq:linforcegali}). Finally, the {\it full model} is, as the name suggests, the Galileon model with its modified background and scale-dependent (with screening) fifth force. Comparing the results of the full and linear variants allows one to measure the impact of the scale dependence of the fifth force, while the QCDM model serves as the reference against which one can measure the effects of the modified gravitational law \footnote{We do not use standard $\Lambda$CDM as the reference model since the latter differs from the Galileon model also in the time evolution of the cosmological background. Here, we are interested on the effects of the fifth forces alone, which is why we use the QCDM variant.}. Similarly, for the Nonlocal model, we also have the equivalent QCDM and full model variants.  For the Nonlocal model, as there is no screening, there is no distinction between the linear and the full models.

We show results from simulation boxes of side $400{\rm Mpc}/h$ for the Galileon, and $200{\rm Mpc}/h$ for the Nonlocal model, both with $512^3$ dark matter tracer particles (these were the boxes used in Refs.~\cite{Barreira:2013eea, Barreira:2014kra}). Each of the model variants was simulated five times using different realizations of the initial density field. We use the variance across the realizations to compute errorbars. When finding voids in the simulations, we shall also make use of DM haloes found in the simulations. Our halo catalogues were obtained with the publicly available {\tt Rockstar} code \cite{Behroozi:2011ju}, which is a phase space {\it friends-of-friends} based halo finder. The number density of the haloes we consider is $n_{\rm halo} = 5\times 10^{-4} h^3/{\rm Mpc}^3$ and $n_{\rm halo} = 5\times 10^{-3} h^3/{\rm Mpc}^3$ for the Galileon and Nonlocal gravity simulations, respectively. This is roughly the number density of haloes after retaining only those haloes whose mass is at least $100$ times the particle mass, $M_p$. This minimum halo mass is  $100M_{p} \approx 4\times 10^{12} M_{\odot}/h$ and $100M_{p} \approx 5\times 10^{11} M_{\odot}/h$, for the simulations of the Cubic Galileon and Nonlocal models, respectively \footnote{Note that due to the different growth of structure, the halo mass function differs between the different variants of the models. The halo catalogues of the different variants were cut at slightly different mass values to yield the same number density of haloes.}. We refer the reader to Refs.~\cite{Barreira:2013eea, Barreira:2014zza, Barreira:2014kra} for further details about the properties of dark matter haloes in these models.

{We find voids using the Watershed Void Finder (WVF) method of Ref.~\cite{Platen:2007qk}. Our code takes as input the discrete tracer distribution, which in our case are DM particles and/or DM haloes, to construct a continuous volume-weighted density field using a Delaunay Tessellation Field Estimator (DTFE) method \cite{Schaap:2000se, 2009LNP...665..291V}. For computational convenience, the DTFE field is sampled onto a regular grid, whose cell size is of the order of the mean distance between tracers. The grid density field is smoothed with a Gaussian filter of size $2\ {\rm Mpc}/h$ to reduce small scale features that could lead to spurious voids \cite{Platen:2007qk}. In the language of the watershed technique, the resulting density field is viewed as a landscape that will be flooded by a rising level of water. The regions around every local minima of the density field are called {\it catchment basins} (where water collects) and will be identified as the voids. As the water level rises, the basins grow and, eventually, neighbouring basins meet at the higher-density ridges that separate them. These ridges mark the boundary of each basin/void, and are associated with the filaments and walls of the cosmic web \citep{2013MNRAS.429.1286C,Cautun:2014fwa}. The process stops when the water level reaches the global maximum of the density field, by the end of which all basin/void boundaries have been identified. To overcome watershed over-segmentation\footnote{This refers to avoid finding too many small voids inside a large underdense region, where in fact the whole underdense region should be classified as a single void that resulted from the merging of smaller ones.}, ridges whose density constrast is $\delta < -0.8$ are not classified as void boundaries, as such low density boundaries are indicative of subvoids that have merged \cite{Sheth:2003py, Platen:2007qk}. An appealing aspect of the watershed method is that it makes no {\it a priori} assumptions on the size, shape or mean underdensity of the voids (see Ref.~\cite{Colberg:2008qg} for a comparison study of different void finders). }

{As is customary in void studies, we define the effective void radius $R_v$ as the radius of a sphere whose volume is the same as the volume of the watershed void. We take the center of the void to be the location of the barycenter which we define as $\vec{r}_{\rm barycenter} = \sum_i\vec{r}_i/N_{\rm cell}$, where $\vec{r}_i$ is the position of each grid cell identified as part of the void and $N_{\rm cell}$ is the total number of grid cells associated with void. We evaluate the density profiles of the voids using the DM density field (for voids found in both the DM and halo density fields) since this is the mass distribution that determines the lensing signal. In what follows we limit ourselves to analysing the simulation results at $z = 0$.}

\subsection{Void size function}\label{sec:counts}

\begin{figure*}
	\centering
	\includegraphics[scale=0.39]{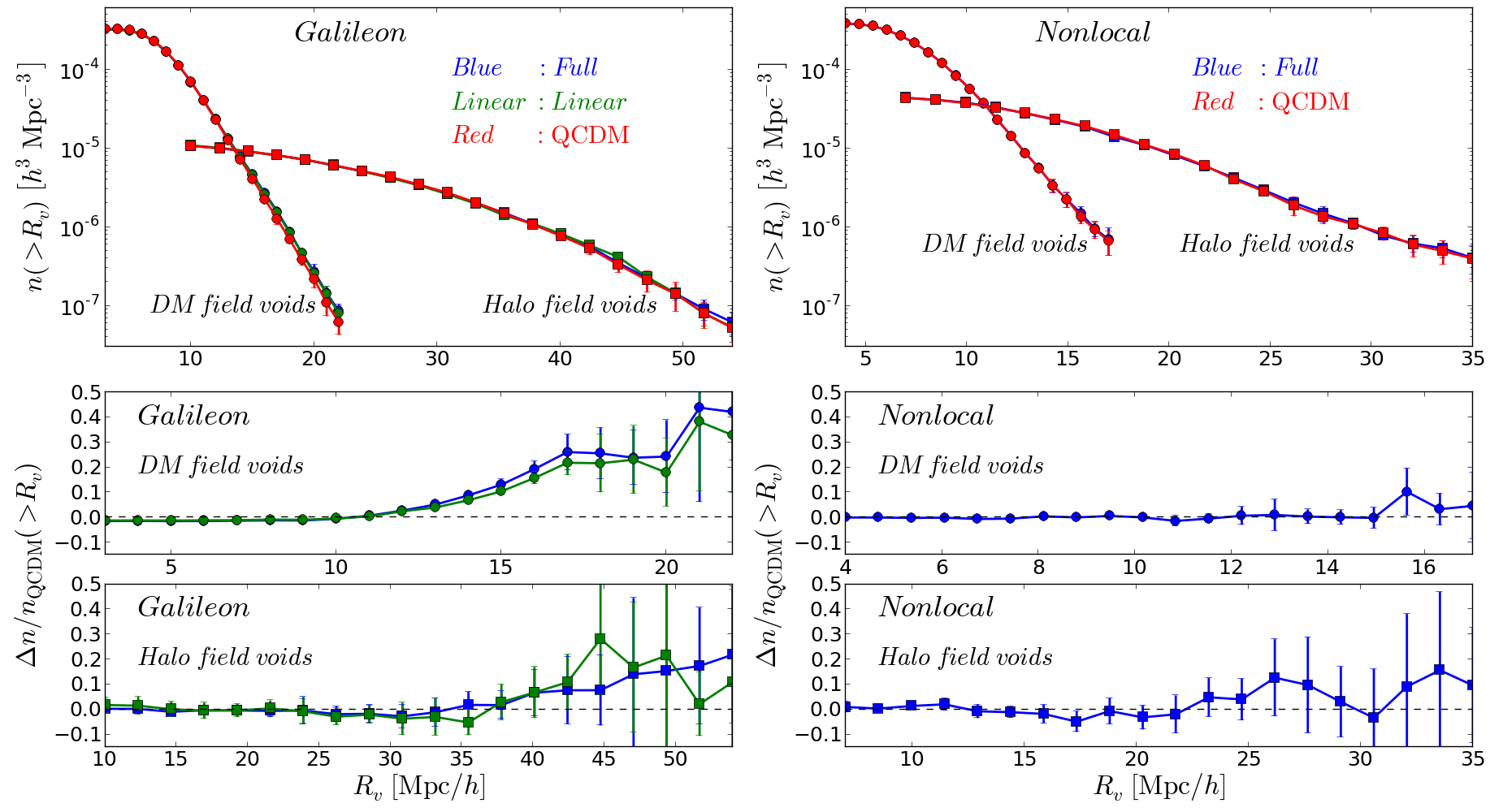}
	\caption{Cumulative void size function (number density of voids with radii above $R_v$) for the Cubic Galileon (left panels) and Nonlocal gravity (right panels) models. The upper panels show the number density of voids found in the DM (circles) and halo (squares) density fields for the full (blue), linear (green) and QCDM (red) variants of each model, as labelled. The lower panels show the difference relative to QCDM. The errorbars depict the variance across the five realizations of each variant.}
\label{fig:counts} \end{figure*}

Figure \ref{fig:counts} shows the cumulative size function of the voids found in the simulations of the Cubic Galileon (left panels) and Nonlocal (right panels) gravity models. For both models, the void population depends on the tracer type used. In particular, DM density field voids (circles) are smaller and, in total, are found in greater number than voids in the halo density field (squares). This follows straightforwardly from the fact that the distribution of collapsed haloes is sparser than that of the DM particles. It is also noteworthy that, for the same type of tracer, we find larger voids in the Cubic Galileon than in the Nonlocal gravity model. Part of this result is due to the fact that the box size used in the simulations of the Galileon model ($400{\rm Mpc}/h$) is larger than that used in the simulations of the Nonlocal model ($200{\rm Mpc}/h$). One should therefore bear this difference in the box size in mind when comparing the results between the two gravity models.

In terms of the relative difference to QCDM, the full and linear variants of the Galileon model predict an enhancement of the order $10\%-20\%$ for the larger DM field voids ($15{\rm Mpc}/h \lesssim R_v \lesssim 20{\rm Mpc}/h$). This is due to the enhanced gravity of these models which boosts the evacuation of matter from inside the voids and the formation of large scale structures. In other words, voids expand faster in the full and linear variants, which is why large voids are more abundant. By the same reasoning, one should also expect the number of smaller voids to be suppressed in the linear and full variants, compared to QCDM. This is because the faster expansion of the voids makes it more likely for small neighbouring voids to merge into larger ones. In Fig.~\ref{fig:counts}, this suppresion can be seen for $R_v \lesssim 10{\rm Mpc}/h$, although to a lesser extent than the enhancement seen for larger DM field voids. Another interesting aspect that is seen in the void abundances of the Galileon model is that the results of the full and linear variants are rather similar. This is very different from what is seen in the abundances of collapsed haloes, for which, due to the suppression effects of the screening mechanism, the full model has considerably fewer massive haloes than the linear variant (see e.g.~Fig.~5 of Ref.~\cite{Barreira:2013eea}.) This illustrates that the effects of the screening mechanism are much weaker around underdense regions, as expected.

In the case of Nonlocal gravity, the number density of DM field voids is, within the errorbars, the same in the full and QCDM variants. Here, recall that the largest voids found in the Nonlocal simulations are smaller than those in the Galileon simulations due to the smaller box size used. For instance, the largest DM field void found in the simulations of the Nonlocal model has $R_v \approx 17 {\rm Mpc}/h$. This, together with the fact that in the Galileon model the enhancement is most noticeable for $R_v \gtrsim 15 {\rm Mpc}/h$, suggests that the simulation box of the Nonlocal gravity model is not big enough to capture the impact of the fifth force on larger voids. Indeed, for large voids, there seems to be a trend for the full Nonlocal model to overpredict the number of voids with $R_v \gtrsim 15 {\rm Mpc}/h$ relative to QCDM, although this is not significant due to the size of the errorbars. Nevertheless, for $R_v \sim 15 {\rm Mpc}/h$, the enhancement in the full and linear variants of the Galileon model is already around $\sim 10\%$, whereas in the Nonlocal model it is still consistent with zero. This shows that the effects of the modifications to gravity in the Nonlocal model are, in general, weaker than those in the Cubic Galileon, which is also expected.

The results become noisier for voids found in the halo field due to the smaller number of tracers. For both the Galileon and Nonlocal gravity models, within the errorbars, the number density of voids is essentially the same in all model variants. However, as an exercise, if one ignores the size of the errorbars for a moment, then one notes that, at least qualitatively, the halo field voids show a similar behaviour to their DM field counterparts. In the case of the Galileon, for instance, the largest halo field voids, $R_v \gtrsim 40{\rm Mpc}/h$, are $\sim 10-20\%$ more abundant in the full and linear variants, compared to QCDM. This qualitative trend, backed up by the expectation based on physical intuition, suggests that with improved halo field void statistics one should recover, at least to a certain degree, the same physical behavior seen for the DM field voids.

In the results that follow, we analyse our void catalogues by splitting them into two bins of radial size. We split the voids in the Galileon model according to
\bq\label{eq:binsgali}
DM\ field\ &::& \ {\rm bin\ 1} \ \left[5-12.5\right]{\rm Mpc}/h,\ \ \ \bar{R}_v \approx    8.20 {\rm Mpc}/h,  \nonumber \\
DM\ field\ &::& \ {\rm bin\ 2} \ \left[12.5-20\right]{\rm Mpc}/h,\ \bar{R}_v \approx       14.2 {\rm Mpc}/h, \nonumber \\
Halo\ field\ &::& \ {\rm bin\ 1} \ \left[10-30\right]{\rm Mpc}/h,\ \ \ \ \bar{R}_v \approx 20.0 {\rm Mpc}/h, \nonumber \\
Halo\ field\ &::& \ {\rm bin\ 2} \ \left[30-50\right]{\rm Mpc}/h,\ \ \ \ \bar{R}_v \approx 36.5 {\rm Mpc}/h, \nonumber \\
\eq
and in the Nonlocal model as
\bq\label{eq:binsnloc}
DM\ field\ \ &::& \ \ {\rm bin\ 1} \ \ \ \left[4-10\right]{\rm Mpc}/h,\ \ \ \bar{R}_v \approx    7.30 {\rm Mpc}/h, \nonumber \\
DM\ field\ \ &::& \ \ {\rm bin\ 2} \ \ \ \left[10-16\right]{\rm Mpc}/h,\ \bar{R}_v \approx       11.4 {\rm Mpc}/h, \nonumber \\
Halo\ field\ \ &::& \ \ {\rm bin\ 1} \ \ \ \left[10-20\right]{\rm Mpc}/h,\  \bar{R}_v \approx    14.5 {\rm Mpc}/h, \nonumber \\
Halo\ field\ \ &::& \ \ {\rm bin\ 2} \ \ \ \left[20-30\right]{\rm Mpc}/h,\ \bar{R}_v \approx     23.2 {\rm Mpc}/h, \nonumber \\
\eq
where $\bar{R}_v = \sum_{i}R_{v,i}/N_{\rm bin}$ is the mean void size in each bin, where $R_{v,i}$ is the radius of the $i$-th void in the bin and $N_{\rm bin}$ is the number of voids in each bin. The exact value of $\bar{R}_v$ fluctuates only slightly ($<1\%$) in between the different model variants. We have found that this binning choice constitutes a good compromise between having enough voids in each bin, whilst making sure that the void properties do not vary too much within a bin.

\subsection{Void density profiles}\label{sec:profiles}

\begin{figure*}
	\centering
	\includegraphics[scale=0.39]{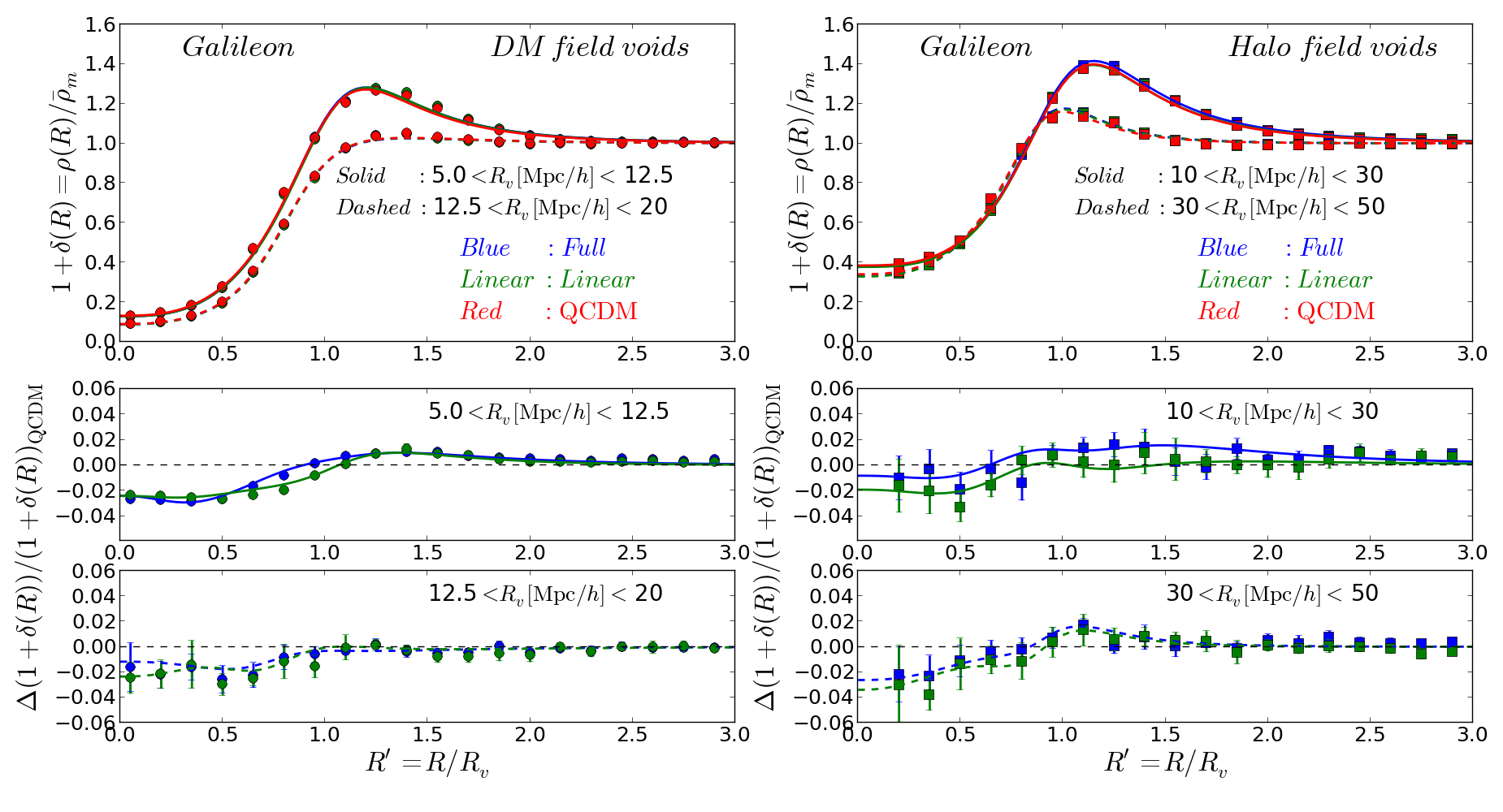}
	\caption{Void density profiles, $\rho_m/\bar{\rho}_m = 1 + \delta$, for the DM density field (circles, left panels) and halo density field (squares, right panels) voids found in the simulations of the three Galileon model variants (distinguished by the different colors, as labelled), plotted as function of the scaled radius $R' = R/R_v$. The profiles are computed using the DM density field (which is the one felt by photons), for both the voids found in the DM and halo density fields. The solid (dashed) lines show the best-fitting density profiles, using the formula of Eq.~(\ref{eq:mine}), for the bin of smaller (larger) void sizes, as labelled. The bottom panels show the relative difference to QCDM. The errorbars depict the variance accross the five realizations of each variant.}
\label{fig:profiles-gali} \end{figure*}

\begin{figure*}
	\centering
	\includegraphics[scale=0.39]{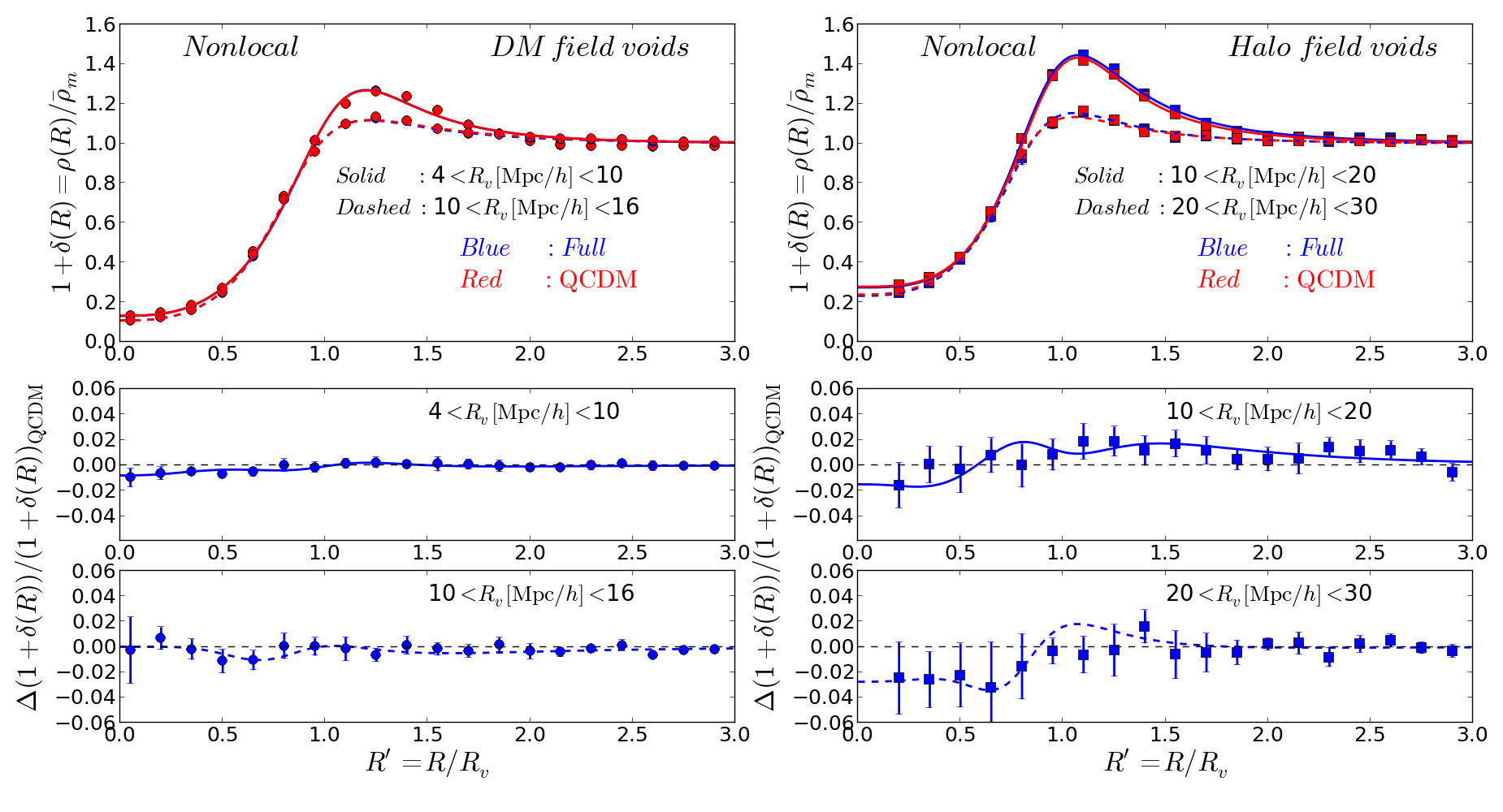}
	\caption{Same as Fig.~\ref{fig:profiles-gali} but for the Nonlocal gravity model.}
\label{fig:profiles-nloc} \end{figure*}

Figure \ref{fig:profiles-gali} shows the spherically averaged DM density field and halo density field void density profiles found in the simulations of the three variants of the Galileon model\footnote{{The average density profile of all the voids in each bin should be spherical to a good approximation, even though each individual void is not.}}, with the void sample split to the size bins according to Eq.~(\ref{eq:binsgali}). Figure \ref{fig:profiles-nloc} is the same as Fig.~\ref{fig:profiles-gali}, but for the Nonlocal model. The void density profiles are characterized by a density increase from $R' = 0$ towards $R' \approx 1$; an overdense {\it ridge} at $R'\sim[1-1.5]$, which is associated with the filaments and walls that surround the void (the ridge is less pronounced for larger voids); and a steady decrease towards the cosmic mean, $\delta = 0$, at larger radii. In these figures, the curves show the best-fitting profiles obtained using the five-parameter formula of Eq.~(\ref{eq:mine}), which fits the simulation results very well. Recall that the density profiles were computed using the DM density field (which is the mass density felt by photons), for both the voids identified using the DM and halo distributions. This fact should be taken into account when comparing these profiles with others in the literature \cite{Hamaus:2014fma, Nadathur:2014qja, Nadathur:2015qua}.

{The functional form of Eq.~(\ref{eq:mine}) is inspired by the expressions proposed by earlier works \cite{Hamaus:2014fma, Nadathur:2014qja} in the context of $\Lambda$CDM. In particular, the formula proposed by Ref.~\cite{Hamaus:2014fma} differs from ours by fixing $s_2 = 1$. This was used to fit to the density profiles of voids found from subsampled (i.e. diluted) DM tracer particle fields in $\lcdm$ (see Ref.~\cite{Hamaus:2014fma} for the details). The authors of Ref.~\cite{Hamaus:2014fma} further found that there are relations between the four free parameters of their formula, which can be used to effectively fix two of them. On the other hand, the formula proposed by Ref.~\cite{Nadathur:2014qja} has $s_2=s_1$, and was used to fit the density profiles of voids constructed from mock and observed galaxy catalogues. In both of these works, the voids were found using watershed-based void finders, as in this paper. Recently, Ref.~\cite{Nadathur:2015qua} explored the connection between the properties of voids found using watershed methods and the predictions of theoretical models based on excursion set theory \cite{Sheth:2003py}. In Ref.~\cite{Nadathur:2015qua}, the authors also pointed out that the performance of the fitting formulae proposed by Refs.~\cite{Hamaus:2014fma, Nadathur:2014qja} may depend on some aspects of the analysis such as the tracer type, tracer density, definition of void center, etc.~(see Ref.~\cite{Nadathur:2015qua} for the details). Compared to these other formulae, our void profile of Eq.~(\ref{eq:mine}) may appear less appealing due to the fact that it has the extra free parameter $s_2$. However, the flexibility that comes with $s_2$ is what allows our formula to be a very good fit to the simulation results, both for the voids found in the DM and halo fields, and for all the variants of the models of gravity we consider. Moreover, the differences relative to QCDM computed using the best-fitting formulae also match very well the relative differences measured in the simulations (lower panels of Figs.~\ref{fig:profiles-gali} and \ref{fig:profiles-nloc}). We note that it is not the goal of this paper to determine if the voids found in our simulations are {\it self-similar} (i.e., independent of the void size) or {\it universal} (i.e., independent of tracer type and/or redshift). It is also not our goal to investigate the dependence of the void profiles on the value of the density at $R = 0$ (in Ref.~\cite{Nadathur:2015qua} the void profiles are shown to depend also on this value, in addition to the $R_v$ dependence). From hereon, in our analysis, what is important to retain is that the void density profiles in the simulations are well described by Eq.~(\ref{eq:mine})\footnote{Even if one needs to fit the free parameters for different void sizes and for different density tracers.}, which can therefore be used to calculate the force profiles and lensing signal.}

The impact of the fifth force is better seen when comparing the difference relative to QCDM. In the case of the Galileon model (Fig.~\ref{fig:profiles-gali}), compared to QCDM, the voids in the full and linear variants are $\approx 2-3\%$ emptier in the inner regions, i.e. $R' \lesssim 0.5$, for both the DM and halo voids (although the result is noisier for halo voids {due to poorer statistics, specially for the smaller radius bin}). Physically, this is because the enhanced gravity favours the piling up of matter in the outer regions, leaving less matter inside the void. The fact that the prediction from the linear and full variants are so close illustrates, once again, that the effects of the screening mechanism are weak around voids. In Ref.~\cite{Cai:2014fma}, similar results were found in the context of $f(R)$ gravity, using a spherical underdensity based void finder \cite{Padilla:2005ea}. In particular, the authors of Ref.~\cite{Cai:2014fma} found that the voids in $f(R)$ models can be up to $\approx 5\%$ emptier than in $\Lambda$CDM. In Fig.~\ref{fig:profiles-gali}, it is also worth noting that for the smaller radius bin of DM field voids, at $R' \sim 0.5 -1$, the voids are more underdense in the linear variant than in the full model. We shall present an explanation for this in the next subsection, when we look at the force profiles in the Galileon model.

The effects of the fifth force on the void profiles of the Nonlocal model (Fig.~\ref{fig:profiles-nloc}) are weaker than those seen in the Galileon case. In particular, for the DM field voids, the smaller void size bin in the full variant shows a decrement of only $\approx 1\%$, relative to QCDM; the difference becomes consistent with zero for the larger size bin. {In the case of the halo field voids, there is a systematic trend for the voids in the full Nonlocal model to be $\approx 2-3\%$ emptier than in QCDM for $R' \lesssim 0.5$, but the poorer statistics make it hard to draw any decisive conclusions.} Nevertheless, the result of Fig.~\ref{fig:profiles-nloc} shows that, overall, the fifth force effects on the void density profiles in the Nonlocal model are weaker than in the Galileon model, which is expected.

\subsection{Force profiles in the Galileon model}\label{sec:forces}

\begin{figure*}
	\centering
	\includegraphics[scale=0.39]{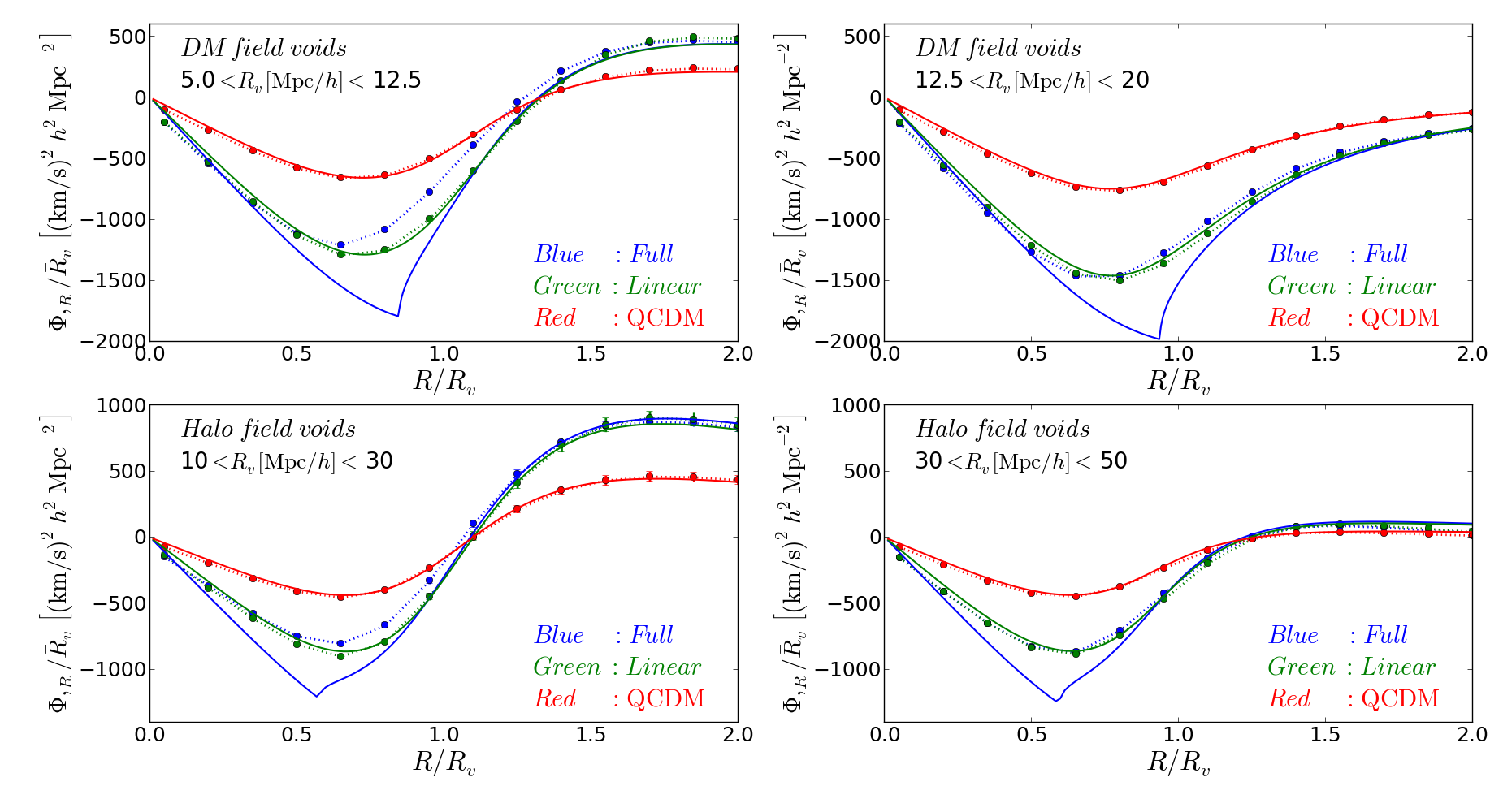}
	\caption{Radial force profiles around the DM field (upper panels) and halo field (lower panels) voids in the variants of the Galileon model (distinguished by the different colors, as labelled). The circles with errorbars (which are in most cases smaller than the circles), linked by the dotted lines, correspond to the spherically averaged radial force field in the simulations. The solid lines correspond to the analytical prediction computed using the corresponding best-fitting void density profiles of Eq.~(\ref{eq:mine}), shown in Fig.~\ref{fig:profiles-gali}. The different panels show the result for the different void size bins, as labelled. What is actually plotted is the radial force scaled by the mean void size in each bin, $\Phi,_R/\bar{R}_v$ (cf.~Eqs.~(\ref{eq:binsgali})). A negative sign for the force means that it points outwards.}
\label{fig:forces} \end{figure*}

Figure \ref{fig:forces} shows the force profiles of the voids in the variants of the Galileon model. The circles, linked by the dotted lines, show the simulation results. These were obtained by spherically averaging the radial force field in the simulations, which was constructed by using the force information at the N-body particle positions. The solid curves show the analytical result computed using the best-fitting void density profiles of Eq.~(\ref{eq:mine}) (cf.~Fig.~\ref{fig:profiles-gali}). Figure \ref{fig:forces} shows that, for the linear and QCDM variants, the analytical calculation is in very good agreement with the simulation results. However, the same is not true for the case of the full variant of the Galileon model. In this case, the analytical result differs from that of the simulations for $R' \lesssim 1.25$, for DM field voids, and for $R' \lesssim 1.0$ for halo field voids. More specifically, for all cases shown, the analytical result of the full variant always predicts a stronger force (more negative) than the linear variant, inside the void. This result was already seen in Sec.~\ref{sec:gali}, when we analysed the behaviour of the regime (iii) discussed there (cf.~Fig.~\ref{fig:complex}). On the other hand, in the simulations, the force inside the smaller voids (left panels) of the full variant is weaker than in the linear case ($R' \sim 0.5-1$). For larger voids (right panel), the full and linear variant simulations exhibit nearly the same force profiles.

The reason why the forces in the simulations of the full model are weaker (less negative) than those computed analytically using the spherically averaged density profiles can be linked to the effects of screening. The smooth void density profiles depicted in Fig.~\ref{fig:profiles-gali} correspond only to an average density field, which does not fully capture the detailed distribution of matter around the voids. A more realistic picture is that, inside the voids and at their edges, there are higher density peaks associated with dark matter haloes and their respective infall regions. Close to these higher density regions, the fifth force in the Galileon model is suppressed by the screening mechanism (cf.~regime (i) discussed in Sec.~\ref{sec:gali}), which results in a weakening of the total force\footnote{{We note also that, in the simulations, the {\it ad-hoc} fix to keep the fifth force a real number is applied on a cell-by-cell basis on the adaptive mesh, which means that close to these density peaks the fix is not employed.}}. Herein lies the explanation for the mismatch between the analytical result and the simulation force profiles. By averaging first the matter field, despite of the presence of higher density peaks, on average, one ends up with a smoother and lower density void profile. This profile, when used in the analytical calculation, gives a fifth force which is stronger in magnitude than the corresponding linear solution (cf.~regime (iii) in Sec.~\ref{sec:gali}). On the other hand, by averaging directly the forces at the particle positions, one is averaging a force field which is already affected by the suppression effects of the screening due to the existing higher density peaks. This is why the force profiles measured in the simulations are weaker (less negative) than the analytical result, as seen in Fig.~\ref{fig:forces}. In other words, since the force equation in the full model is nonlinear (cf.~Eqs.~(\ref{eq:phiror})), it makes a difference whether one computes the force analytically from the averaged density field, or one computes the force by averaging directly the force field. In the case of the QCDM and linear variants, the force equation is linear, and as a result, the operations of averaging the density and the force field commute, which is why there is almost perfect agreement between the analytical and simulation results in these cases.

Figure \ref{fig:forces} shows also that the suppression of the total force in the simulations of the full variant relative to the linear one is more pronounced in smaller voids. This is because smaller voids are denser, and therefore contain more higher density peaks per volume inside them and in their surroundings, which enhances the suppresion effect of the screening. In particular, it is interesting to link this result with the differences between the linear and full model density profiles for the smaller size bin of the DM field haloes  at $R' \sim 0.5-1$ in Fig.~\ref{fig:profiles-gali}. As we noted in the previous section, on these radial scales the voids in the linear model are slightly emptier. This can be explained by the fact that, in the simulations, the force in the linear model is stronger (more negative), which favours the evacuation of matter from inside the void\footnote{It is worth noting that the relative differences in Fig.~\ref{fig:profiles-gali} correspond to different void populations, and as a result, some of the observed differences could arise from this. As a test, we have measured the density and force profiles in the full, linear and QCDM simulations, but at the spatial locations of the voids in the QCDM model. This increases the chances of comparing voids that evolved from the same initial underdense regions. From this test we found only small quantitative changes with no impact on our conclusions.}. In principle, the same result should also be noticeable in the case of the smaller size bin of the halo field voids, for which the force in the full model is also weaker than in the linear one at $R'\sim0.5-1$ (lower left panel of Fig.~\ref{fig:forces}). This is not visible in Fig.~\ref{fig:profiles-gali} (lower left panel), possibly because of the noisier measurements.

For the Nonlocal gravity model, the force equations are linear, as in the QCDM and linear variants of the Cubic Galileon model. As a result, there is good agreement between force profiles computed from simulations and the analytical results.


\section{Weak Lensing by voids in modified gravity}\label{sec:lensing}

In this section, we analyse the gravitational lensing signal from the voids in the Cubic Galileon and Nonlocal gravity models. We start by describing how to calculate the relevant lensing quantities in these theories of gravity, and then focus on the predictions using the voids found in the simulations of the two models.

\subsection{Lensing shear calculation}\label{sec:calculation}

\subsubsection{Lensing shear in GR}\label{sec:grlens}

The observable quantity in weak lensing studies \cite{Bartelmann:1999yn, 2010CQGra..27w3001B} is the {\it reduced shear}, $g = \gamma_t/(1 - \kappa)$, where $\gamma_t$ and $\kappa$ are called, respectively, the lensing tangential shear and convergence. The reduced shear is directly related to the ellipticities of the background galaxies whose light is distorted by intervening gravitational sources, which are voids in our case. In the weak-lensing regime, which is the regime for voids \cite{Amendola:1998xu, 2013ApJ...762L..20K},  $\gamma_t$ and $\kappa$ are both much smaller than unity, and consequently, one has $g \approx \gamma_t$. 

The lensing convergence is obtained by integrating the Laplacian of the lensing potential $\Phi_{\rm len} = \left(\Phi + \Psi\right)/2$ as
\bq\label{eq:k-def}
\kappa = \frac{1}{4\pi G \Sigma_c} \int \nabla^2\Phi_{\rm len} {\rm d}l,
\eq
where $l$ is the line of sight coordinate. Here, $\Sigma_c = D_sc^2/(4\pi G D_{ds}D_d)$ is called the critical surface mass density for lensing, where $D_d$, $D_s$ and $D_{ds}$, are respectively, the angular diameter distances between the observer and the void,  the observer and the source galaxies, and the void and the source galaxies. We note, however, that given the way we choose to present our results below, the exact values of $\Sigma_c$ are not important (we comment further on this below). In GR, the two Newtonian potentials are the same $\Phi = \Psi \equiv \Phi_{\rm len}$ (in the absence of anisotropic stress). As a result, using the Poisson equation $\nabla^2\Phi = 4\pi G\bar{\rho}_m\delta$, the convergence is given by
\bq\label{eq:k-def-2}
\kappa = \frac{\bar{\rho}_m}{\Sigma_c} R_v \int \delta(r', l') {\rm d}l' \equiv R_v\frac{\Sigma(r')}{\Sigma_c},
\eq
where $r' = r/R_v$, $l' = l/R_v$, and $r$ is a two-dimensional radial coordinate defined on the void plane ($l = 0$) with origin at the void center (i.e.~$R^2 = r^2 + l^2$). From the above equation, one sees that in GR the lensing convergence is simply given by the projected density profile of the void, $\Sigma(r) = R_v\Sigma(r')$. We perform the integral of Eq.~(\ref{eq:k-def-2}) numerically, using the density contrast formula of Eq.~(\ref{eq:mine}). Note that the integral is dimensionless, and hence, $\kappa$ is dimensionless as well.

The tangential shear is defined as
\bq\label{eq:shear}
\gamma_t  = {\overline{\kappa} - \kappa},
\eq
where
\bq\label{eq:meank}
\overline{\kappa} = \frac{2}{r^2}\int_0^r y \kappa(y) {\rm d}y \equiv R_v \frac{2}{r'^2}\int_0^{r'} y' \kappa(y') {\rm d}y'
\eq
is the mean convergence inside radius $r$. The mean projected mass inside radius $r$ is given by $\overline{\Sigma}(<r) = \overline{\kappa}\Sigma_c$. For clarity, we note that in the last equality of Eq.~\ref{eq:meank} we have used that  $\kappa(y) = R_v\kappa(y')$, where $\kappa(y')$ is the convergence given in terms of the radial coordinate scaled by $R_v^{-1}$. Here, we follow Refs.~\cite{Melchior:2013gxd, Clampitt:2014gpa} and quote the lensing predictions in terms of the {\it differential surface mass density} $\Delta\Sigma(r)$, which is given by
\bq\label{eq:dmd-gr}
\Delta\Sigma(r) = \overline{\Sigma}(<r) - \Sigma(r) = \Sigma_c\gamma_t.
\eq
By quoting the results in terms of $\Delta\Sigma(r)$ we avoid having to compute the values of $\Sigma_c$. In Eqs.~(\ref{eq:k-def-2}) and (\ref{eq:meank}), we have also factored out the void radius, $R_v$, to make it explicit that the lensing quantities, depend linearly on it, e.g.~$\Delta\Sigma \propto R_v$.

\subsubsection{Lensing in linear models of gravity with $G_{\rm eff}(z)$}\label{sec:gefflens}

As we have seen in Secs.~\ref{sec:nloc} and \ref{sec:gali}, in Nonlocal gravity and in the linear variant of the Cubic Galileon model, the gravitational law is the same as in GR, but with a scale-independent effective gravitational strength, $G_{\rm eff}$. Moreover, in these two models, the two Newtonian potentials are also the same and equal to the lensing potential, $\Phi_{\rm len} = \Phi = \Psi$ (in the absence of anisotropic stress). Consequently, the differential surface mass density in these models is obtained in the same way as in GR, but taking the factor $G_{\rm eff}/G$ properly into account. Explicitly, one has
\bq\label{eq:dmd-lin}
\Delta\Sigma(r) = \frac{G_{\rm eff}}{G} \Delta\Sigma^{\rm GR}(r).
\eq

\subsubsection{Lensing in the full Cubic Galileon model}\label{sec:fulllens}

As in GR, in the full variant of the Galileon model one also has that $\Phi_{\rm len} = \Phi = \Psi$, and therefore, the lensing convergence is also given by integrating $\nabla^2\Phi$ along the line of sight

\bq\label{eq:k-full}
\kappa = \frac{1}{4\pi G\Sigma_c} R_v \int \left(\Phi,_{RR}(r', l') + 2\frac{\Phi,_R}{R}(r', l')\right) {\rm d}l', \nonumber \\
\eq
where $\Phi,_R/R$ and $\Phi,_{RR}$ are given, respectively, by Eqs.~(\ref{eq:phiror}) and (\ref{eq:phirr}), and recall that in spherical coordinates, $\nabla^2\Phi = \Phi,_{RR} + 2\Phi,_R/R$. Given $\kappa$,  then the values of $\overline{\kappa}$ and $\gamma_t$ are obtained as in Eqs.~(\ref{eq:meank}) and (\ref{eq:shear}), respectively. Here, we shall also quote the results for the full Galileon model in terms of the differential surface mass density, $\Delta\Sigma = \Sigma_c\gamma_t$. However, one should bear in mind that the meaning of $\Delta\Sigma$ is different from the previous cases. In the full Galileon model, in addition to the contribution from the projected mass, $\Delta\Sigma$ also depends on the projected distribution of the Galileon field (cf.~Eqs.~(\ref{eq:phiror}) and (\ref{eq:phirr})).

\subsection{Lensing by the voids in the simulations}\label{sec:basics}

\begin{figure*}
	\centering
	\includegraphics[scale=0.39]{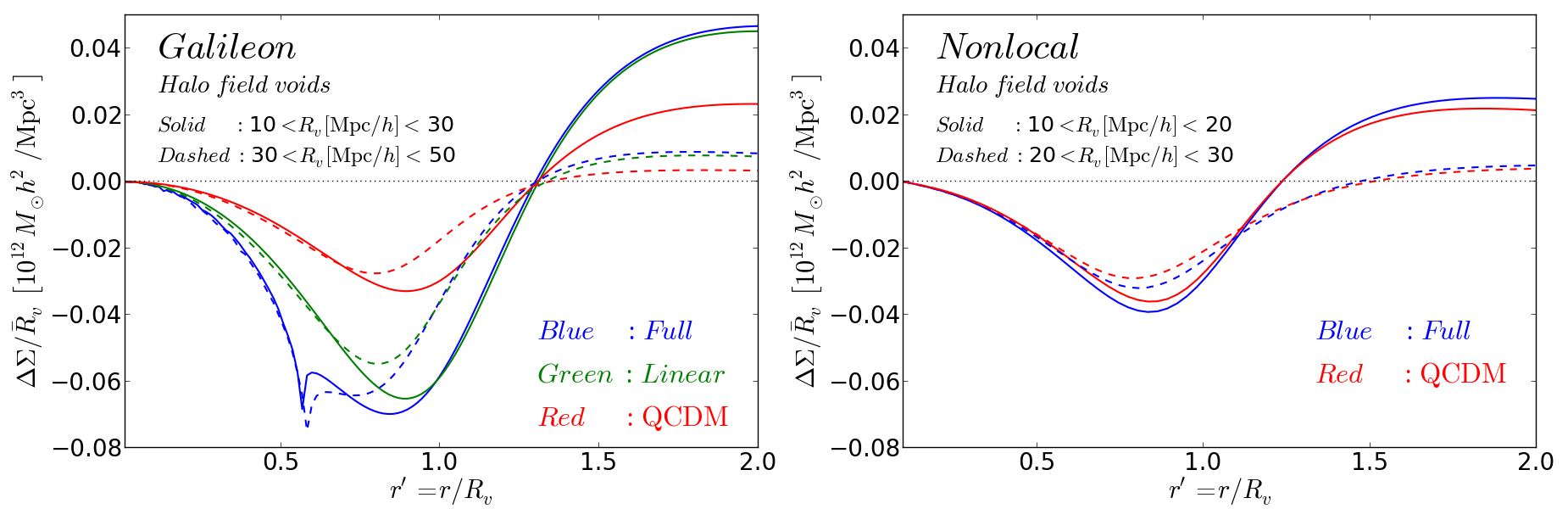}
	\caption{Lensing differential surface mass density, $\Delta\Sigma$, for the best-fitting density profiles of the two void size bins (distinguished by the line styles, as labelled) for the halo density field voids found in the simulations of the different variants (distinguished by the colors, as labelled) of the Cubic Galileon (left panel) and Nonlocal (right panel) gravity models. The result is scaled by the inverse of the mean void radius in each size bin, $\bar{R}_v^{-1}$.}
\label{fig:lens} \end{figure*}

\begin{figure}
	\centering
	\includegraphics[scale=0.345]{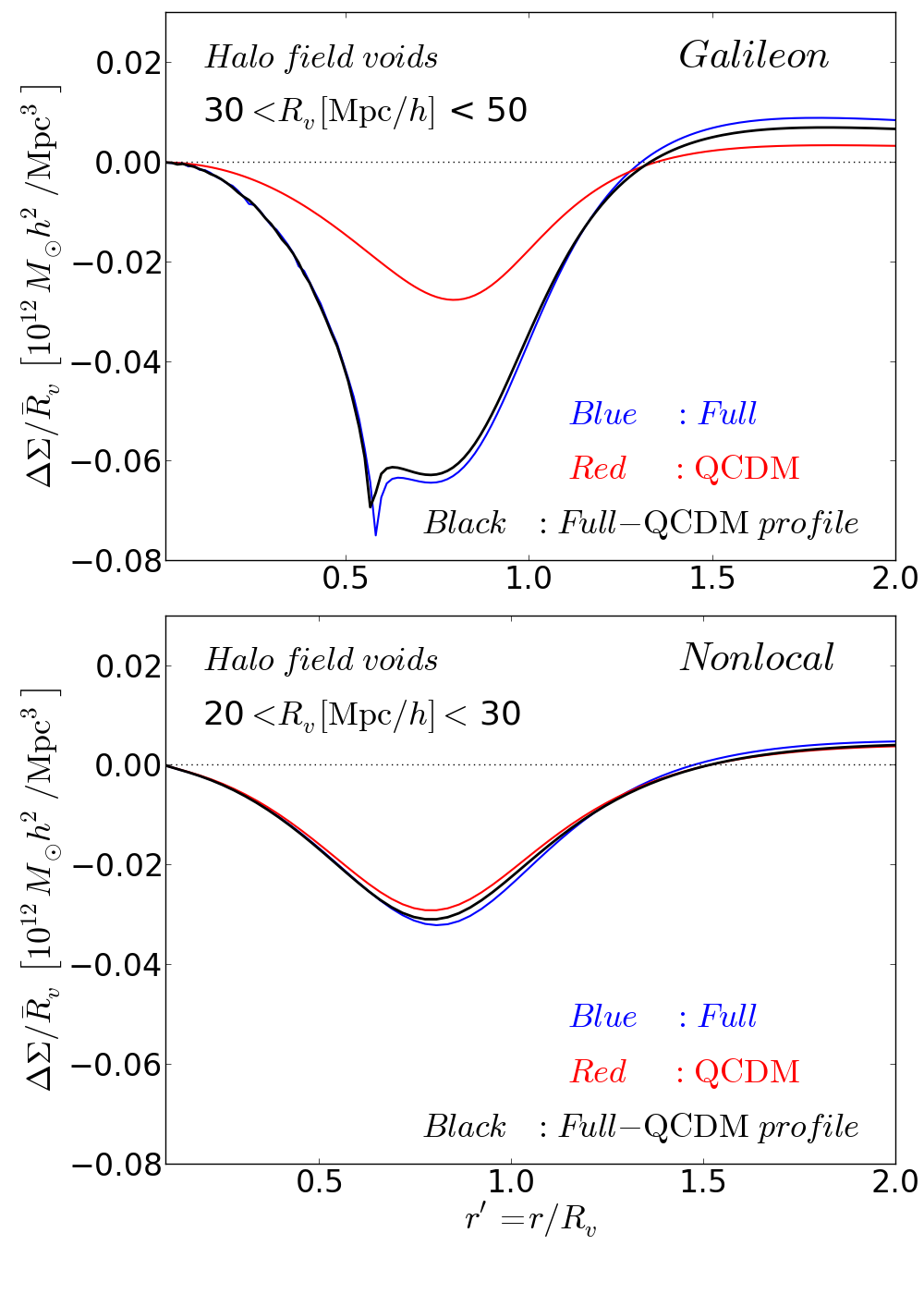}
	\caption{Relative impact of the fifth force and modified density profiles on the lensing signal of halo field voids in the Cubic Galileon (upper panel) and Nonlocal (lower panel) gravity models. For both models, the red and blue lines have the same meaning as those in Fig.~\ref{fig:lens}. The black curves are obtained by calculating the lensing signal with the full Galileon and full Nonlocal model force equations, but using the best-fitting void density profile of the QCDM voids. The comparison between the red and black measures the effects of the fifth force alone; whereas comparing the black and blue curves shows the impact of the modified void density profiles.}
\label{fig:impact} \end{figure}

Figure \ref{fig:lens} shows the lensing signal associated with the halo field voids found in the simulations of the Galileon (left panel) and Nonlocal (right panel) gravity models. The curves were computed as described in the previous subsection using the best-fitting density profiles of Eq.~(\ref{eq:mine}) shown in the right panels of Figs.~\ref{fig:profiles-gali} and \ref{fig:profiles-nloc}. For brevity, we show only the result for halo field voids. These are the ones that are more closely related to observations, where one first identifies voids using galaxy catalogues and then looks at the lensing signal at the void locations \cite{Melchior:2013gxd, Clampitt:2014gpa}. Note also that the values of $\Delta\Sigma$ are scaled by $\bar{R}_v^{-1}$, which means that the voids in the larger size bin (dashed curves) have lensing effects of larger magnitude (i.e.~one needs to multiply the result by $\bar{R}_v$).

In the case of the Galileon model, the maximum of the lensing signal, which occurs at $r' \sim 0.75-1$, is approximately twice as strong in the full and linear variants, compared to QCDM. At $r' \lesssim 1$, the signal is slighlty stronger in the full than in the linear variant \footnote{The "spikey" features seen in the blue curves follow from our {\it ad hoc} fix to the problem of complex fifth force solutions (cf.~Sec.~\ref{sec:gali}).}. This follows from the stronger fifth force (more negative) in the full model associated with the regime (iii) discussed in Sec.~\ref{sec:gali}. However, in Fig.~\ref{fig:forces}, we have seen that, due to the screening mechanism, the force predictions of the full and linear variants measured in the simulations are actually closer than what is predicted by the analytical calculation using the best-fitting density profiles. For this reason, it is reasonable to assume that the lensing signal of the full model is actually closer to the prediction of the linear variant, compared to what is observed in Fig.~\ref{fig:lens}. Nevertheless, given that the differences between the full and linear variants are smaller than their differences relative to QCDM, it remains safe to conclude, as we have seen in previous sections, that the effects of the fifth force in the Galileon model are quite pronounced in voids, where the screening is not very efficient.

In the case of the Nonlocal model, the effects of the fifth force are considerably weaker than in the Galileon model. In particular, for the cases shown in Fig.~\ref{fig:forces}, the maximum amplitude in the value of $|\Delta\Sigma|$ (at $r'\sim0.75-1$) is $\approx 10\%$ larger in the full Nonlocal model, compared to its QCDM variant. This illustrates that the effect of the modifications to gravity in the Nonlocal model are more challenging to detect using the lensing signal from voids.

The modifications to gravity affect the lensing in voids in two main ways: (i) through the modifications to the average density profiles of voids; and (ii) directly through the modifications to the lensing potential. Figure \ref{fig:impact} measures the relative impact of these two effects in the Galileon (upper panel) and Nonlocal gravity (lower panel). In the figure, the red and blue curves are the same as in Fig.~\ref{fig:lens}. The black curves are computed using the lensing equations of the full variants of the Galileon and Nonlocal models, but using the density profile of the voids in the QCDM variants. As a result, comparing the red and black curves shows the effect of modifying the force law, whereas the difference between the black and blue lines shows the impact of the modified density profiles. In the case of the Galileon model, Fig.~\ref{fig:impact} shows that the dominant effect comes from the fifth force. This is seen by the large difference between the red and black curves. Figure \ref{fig:impact} shows the result for the larger size bin of the halo field voids, which are slightly emptier in the full variant of the Galileon model, compared to QCDM (cf.~bottom right panel in Fig.~\ref{fig:profiles-gali}). This helps to further increase the amplitude of $|\Delta\Sigma|$, but by a much smaller amount. On the other hand, for Nonlocal gravity, the direct effect of the fifth force on lensing is comparable to the effect of having slightly emptier voids (cf.~Fig.~\ref{fig:profiles-nloc}). 

\subsection{Connecting to observations}\label{sec:obs}

References \cite{Melchior:2013gxd} and \cite{Clampitt:2014gpa} have recently detected the lensing signal associated with voids in the galaxy distribution (see also Refs.~\cite{Amendola:1998xu, 2013ApJ...762L..20K, 2013MNRAS.432.1021H} for earlier forecast studies). In Ref.~\cite{Melchior:2013gxd}, the authors stacked the voids of the catalogue of Ref.~\cite{2012ApJ...761...44S}, which were found using a watershed algorithm in the three-dimensional main galaxy and luminous red galaxy (LRG) samples of the Sloan Digital Sky Survey-Data Release 7 \cite{Abazajian:2008wr} (SDSS-DR7). On the other hand, in Ref.~\cite{Clampitt:2014gpa}, the authors used also the SDSS LRG catalogues, but the voids were found using a method that is specifically designed for lensing. In this method, emptier regions are found in projected two-dimensional slices of the survey volume, which seems to increase the significance of the lensing detection.

As we discuss below, a robust comparison between these observations and the results of Fig.~\ref{fig:lens} requires more detailed modelling of the theoretical predictions. Nevertheless, one can still compare some of our results to try to get a feeling about what these measurements imply for modified gravity. For instance, in Fig.~5 of Ref.~\cite{Clampitt:2014gpa}, it is shown that for voids with size $ R_v \in \left[15, 30\right]\ {\rm Mpc}/h$, the values of the differential surface mass density at its minimum are, approximately, within $\Delta\Sigma \in \left[-0.4, -0.7\right]\ 10^{12}M_{\odot}h/{\rm Mpc}^2$ (this estimate is based on the size of the errorbars there). From Fig.~\ref{fig:lens}, for the case of the smaller size bin of the full Galileon model we have ${\rm min}\left(\Delta\Sigma\right) \approx -0.065\bar{R}_v = -1.3\times10^{12}M_{\odot}h/{\rm Mpc}^2$. For the full Nonlocal model, we have ${\rm min}\left(\Delta\Sigma\right) \approx -0.032\bar{R}_v = -0.46\times10^{12}M_{\odot}h/{\rm Mpc}^2$ and ${\rm min}\left(\Delta\Sigma\right) \approx -0.039\bar{R}_v = -0.9\times10^{12}M_{\odot}h/{\rm Mpc}^2$, for the smaller and larger size bins, respectively. Hence, for both the Galileon and Nonlocal models of gravity, we get the same typical order of magnitude as in the observations. One notes that in the case of the Galileon model, the size of the effect is larger than the results presented in Ref.~\cite{Clampitt:2014gpa}. This suggests that, indeed, lensing by voids may have the potential to constrain models like the Galileon. 

Before summarizing our results in the next section, we find it instructive to briefly comment on a number of aspects that should to be taken into account before properly confronting these (and other) models to lensing observations. These aspects include:

\setlength{\leftmargini}{10pt}\begin{enumerate}

\item \underline{{\it Impact of $\Sigma_c$}} In Fig.~\ref{fig:lens}, we quote our results in terms of $\Delta\Sigma$, but in reality, what one measures directly from galaxy ellipticities is the shear, $g \approx \gamma_t = \Delta\Sigma/\Sigma_c$. The calculation of $\Sigma_c$ depends on the cosmological background, which can be different between the Galileon, Nonlocal, and the standard $\Lambda$CDM models. Consequently, if in observational studies, one measures $\gamma_t$, but quotes the results in terms of $\Delta\Sigma$ by assuming a background cosmology to compute $\Sigma_c$, then this may introduce some bias that should be carefully addressed. Furthermore, $\Sigma_c$ depends also on the redshift distribution of the source galaxy population, although this can always be set accordingly using the properties of the observed galaxies.

\item \underline{{\it  Void redshift distribution}} The lensing signal in Fig.~\ref{fig:lens} was obtained analytically using the density profiles of the voids in the simulations at $z=0$. In the observations, however, the lensing signal is detected by stacking voids that span a given redshift distribution $z > 0$. In the particular case of the Galileon and Nonlocal gravity models, the fifth force is weaker at earlier times (see e.g.~Fig.~3 of Ref.~\cite{Barreira:2013eea} and Fig.~2 fo Ref.~\cite{Barreira:2014kra}), which reduces the amplitude of the signal depicted in Fig.~\ref{fig:lens}. 

\item \underline{{\it  Void stacking}} The lensing signal associated with individual voids is too weak to be detected in current observations, which is why Refs.~\cite{Melchior:2013gxd, Clampitt:2014gpa} used stacked voids in their analyses. When interpreting such results in modified gravity, for a given stack, voids at different redshifts have different weights in the observed lensing signal because of the redshift dependence of the fifth force, $\Sigma_c$ and also of the screening efficiency. Such effects should be taken into account if one, for instance, tries to use the lensing observations to reconstruct a mean density profile for the stack. Here, an interesting analysis would involve stacks of voids binned not only by size, but also by redshift.

\item \underline{{\it Systematic biases}} {The lensing calculations performed here assume that the density distribution in voids is perfectly smooth. In reality, however, voids contain substructure and its amount is expected to be different in models with different growth rates of structure. Given that the lensing signal from voids is relatively weak (compared to that induced by DM haloes) it may be interesting to investigate the extent to which void substructure can impact on the overall lensing signal. This can be studied by looking at the lensing signal using ray-tracing methods in the simulations, without modelling their profiles as a smooth distribution. Our lensing calculations also assume that the void is the only source of lensing. A ray-tracing analysis would also help to better quantify the contamination of the lensing signal coming from intervening matter along the line of sight.}

\item \underline{{\it  Screening effects}} Related to the above point, a ray-tracing analysis is also able to capture more accurately the effects of the nonlinear screening mechanism. In Sec.~\ref{sec:forces}, we saw that, in the full variant of the Galileon model, it makes a difference whether one computes the force profiles analytically from the spherically averaged density profiles, or by spherically averaging the force field directly. {Moreover, the efficiency of the Vainshtein mechanism depends also on the geometry of the mass distribution as investigated recently in Refs.~\cite{Falck:2014jwa, Falck:2015rsa}. This means that calculations based on the mean spherical profile of a stack of voids may not fully capture the fifth force effects from each individual nonspherical void.} These issues can be circunvented by directly integrating $\nabla^2\Phi$ along the line of sight for each void using ray-tracing and stacking the resulting signal. In this way, one probes directly the lensing potential distribution without introducing any bias that arises when one averages first the density field.

\item \underline{{\it Halo/galaxy abundance and bias}} We have found voids using both DM particles and DM haloes as tracers, with the latter case being that which is more relevant when comparing to observations. Due to halo bias, haloes cluster differently depending on their mass, and hence, the resulting void catalogues depend on the minimum halo mass cut used to identify them. This is turn has an impact on the abundances and profiles of voids \cite{Cai:2014fma, 2015arXiv150608831P, 2015arXiv150700197N}. Since different types of galaxies populate haloes differently, a robust comparison with observations should ensure that the number density and bias of the tracers used in simulations matches those of the tracers used in observations. A first step towards this goal could be to construct mock galaxy catalogues using halo occupation distribution modelling \cite{2004ApJ...609...35K}, as Ref.~\cite{2015arXiv150700197N} has done for $\lcdm$. Such an analysis can also tell whether voids identified using certain types of density tracers are better suited for tests of modified gravity.

\item \underline{{\it Combining different void finders}} The way voids are found in simulations and/or in real galaxy catalogues can also affect the resulting lensing signal. For instance, as we mentioned above, the authors of Ref.~\cite{Clampitt:2014gpa} optimize their analysis for lensing by finding the voids in projected two-dimensional slices of a spectroscopic galaxy survey. This may partly explain why the significance of their detection is higher than that found in Ref.~\cite{Melchior:2013gxd}, in which the voids are found in three dimensions. It would therefore be of great interest to find voids in the way of Ref.~\cite{Clampitt:2014gpa} in the N-body simulations as well. It is also well known that different void finding techniques yield different void profiles \cite{Colberg:2008qg}, and hence, different lensing predictions. For instance, voids found with spherical underdensity (SU) methods have sharper transitions from the inside of the void to the surrounding ridge. This boosts the lensing effect, as can be checked in the $\alpha$ panel of Fig.~\ref{fig:voidparamsimpact} in the Appendix, where $\alpha$ is the parameter of the formula of Eq.~(\ref{eq:mine}) that controls the slope of this transition. Moreover, the recent work of Ref.~\cite{Cautun:2015bma} has shown that it may be more natural to characterize the void profiles with respect to their boundaries (which is where most of the mass is), instead of with respect to the void center (which is devoid of tracers). This also results in steeper density profiles close to the void edge (see Fig.~7 of Ref.~\cite{Cautun:2015bma}). The differences in void profiles obtained with different void finding methods is generally portrayed as a source of uncertainty in void related work, but we note that some advantages may arise from it. For instance, since the fifth force acts to make voids emptier and the ridges denser, then methods like SU or that of Ref.~\cite{Cautun:2015bma} may be particularly suitable for modified gravity studies, as they may amplify the size of the fifth force effects (see e.g.~Ref.~\cite{Cai:2014fma}, where the authors use SU methods to study voids in $f(R)$ gravity). Hence, we believe that the combination of the results from different void finding methods (provided they are consistently used in simulated and real data) is something to be explored with more detail when designing observational tests. These investigations are the subject of ongoing work \cite{mariusprep}.

\end{enumerate}


\section{Summary \& Conclusions}\label{sec:summary}

We have studied the lensing signal associated with voids in Cubic Galileon and Nonlocal gravity cosmologies, which are examples of models that modify the gravitational lensing potential. The gravitational law in the Nonlocal model can be parametrized by an enhanced effective gravitational strength ($G_{\rm eff} \approx 1.06G$, at $z = 0$), which is independent of the length scale. In the Galileon model, the modifications to gravity are scale-dependent and in Sec.\ref{sec:gali} we discussed three relevant regimes: (i) a regime which occurs close to massive bodies, in which the fifth force is suppressed via the Vainshtein screening mechanism; (ii) a regime which occurs in regions of small density constrast, $|\delta| \ll 1$, where the equations become linear and the total force can be parametrized by an effective gravitational strength, ($G_{\rm eff} \approx 2G$, at $z = 0$); and finally, (iii) a regime which occurs in regions where the density contrast becomes sufficiently negative, where the amplitude of the fifth force is the largest (more negative).

The fifth force in the Galileon and Nonlocal gravity models has an impact on the lensing signal in and around voids through two main effects. First, the fifth force changes the density profiles of the voids, and second, it also modifies the lensing potential directly. This means that even for fixed mass distribution, the lensing signal in these theories of gravity is still modified w.r.t.~GR. This is different from other popular models like $f(R)$ and/or DGP gravity, which practically do not directly modify the lensing potential. Hence, models that directly modify lensing are more amenable to being tested by lensing observations than those that do not.

We have used results from N-body simulations to study the effect of the fifth force in these two theories. We analysed the abundances and profiles of the density, force and lensing shear of the voids found in the DM and halo density fields of the simulations using a watershed algorithm. When assessing the impact of the modifications to gravity in these two models, we always compared their predictions to models called QCDM, which have the same background expansion as the respective Galileon and Nonlocal models, but have GR as the theory of gravity. In the case of the Galileon model, we have also analysed the results of a model variant with a linearized scale-independent force law (cf.~Table \ref{table-variants}). Our main results can be summarized as follows:

\hspace{0.2 cm} $\bullet$ In the Galileon model, the fifth force boosts the abundance of the larger radius DM field voids ($R_v \gtrsim 15 {\rm Mpc}/h$) by $\approx 10\%-30\%$ (cf.~Fig.~\ref{fig:counts}). This is because the enhanced gravity causes voids to expand faster and also favours the merging of smaller voids into larger ones. For the voids found in the halo density field, the same qualitative trend is also seen but is less pronounced due to poorer statistics. In the case of the Nonlocal model, the modifications to gravity are not strong enough to leave a clear signal on the abundances of the voids found in our simulations (cf.~Fig.~\ref{fig:counts}).

\hspace{0.2 cm} $\bullet$ Our five-parameter formula of Eq.~(\ref{eq:mine}) fits very well the DM and halo field void density profiles in the simulations for all the model variants and for a wide range of void sizes (cf.~Figs.~\ref{fig:profiles-gali} and \ref{fig:profiles-nloc}). Our formula contains an extra parameter compared to others used recently in the literature \cite{Hamaus:2014fma, Nadathur:2014qja}, which gives it the flexibility required to provide good fits. We used our best-fitting formula to compute analytically the lensing signal associated with the voids in the simulations. {Investigations about the self-similar or universal nature of the profiles in our simulations are left for future work.}

\hspace{0.2 cm} $\bullet$ The fifth force in the Galileon model makes the voids slightly emptier ($\approx 2\%-3\%$) in their inner parts, $R' \lesssim 0.5$ (cf.~Fig.~\ref{fig:profiles-gali}). This is because the enhanced gravitational strength favours the evacuation of matter from the inside of the void into the surrounding filament and wall structures. This result is seen for both DM and halo field voids, although the signal is more significant for the smaller DM field voids. In the case of the Nonlocal model, the gravitational strength is also enhanced, and so one expects the same qualitative behavior. Quantitatively though, the weaker fifth force in this model, together with the size of the errorbars allowed by our simulations, makes it more difficult to see the effects of the modifications to gravity (cf.~Fig.~\ref{fig:profiles-nloc}).

\hspace{0.2 cm} $\bullet$ Inside the voids of the full Galileon model, the force measured directly from the simulations is weaker (less negative) than the force computed analytically from the best-fitting void density profile (cf.~Fig.~\ref{fig:forces}). We have attributed this to the screening by high density peaks that exist inside the voids and in their surroundings, whose effect gets diluted if one averages the density field first to compute the force analytically. On the other hand, for the linear and QCDM variants, the analytical result is in very good agreement with the force measured directly from the simulations.

\hspace{0.2 cm} $\bullet$ The effects of the fifth force in the Galileon model can make the lensing signal in voids approximately twice as strong as in GR (cf.~Fig.~\ref{fig:lens}). This large difference comes predominantly from the modifications of the lensing potential {\it per se}, with the different void density profiles being of secondary importance (cf.~Fig.~\ref{fig:impact}). In the case of the Nonlocal gravity model, the fifth force also enhances the expected suppression of the lensing signal, but only by $\approx 10\%$ (cf.~Fig.~\ref{fig:lens}). In this model, the modifications to the density profiles and direct effects of the fifth force on lensing contribute equally to the difference relative to GR (cf.~Fig.~\ref{fig:impact}).

\hspace{0.2 cm} $\bullet$ For all our Galileon model results, the predictions from the full and linear variants are of comparable size. This is different from the case of predictions associated with dark matter haloes (like their abundances or concentration), for which the effects of the full variant are typically much smaller than those of the linear variant because of the screening \cite{Barreira:2014zza}. This illustrates that the suppression effects of the screening mechanism are not very strong around voids, which is why the latter can be regarded as potentially powerful probes of gravity on cosmological scales.


\bigskip

\bigskip

Overall, the results in this paper show that observations of the lensing signal associated with voids can prove very valuable in constraining the gravitational law on large scales. In the future, we plan to use some of the results presented here to help to develop more robust observational tests, by following the steps outlined in Sec.~\ref{sec:obs}. We believe that such investigations would be timely, specially when interpreted in light of future observational missions such as DESI \cite{Levi:2013gra}, LSST \cite{Abell:2009aa} and Euclid \cite{Laureijs2011}.

\begin{acknowledgments}

We thank Richard Brito, Yan-Chuan Cai, Kazuya Koyama, Seshadri Nadathur and Hans Winther for useful comments and discussions. We also thank Lydia Heck for invaluable numerical support. This work was supported by the Science and Technology Facilities Council [grant number ST/L00075X/1]. This work used the DiRAC Data Centric system at Durham University, operated by the Institute for Computational Cosmology on behalf of the STFC DiRAC HPC Facility (www.dirac.ac.uk). This equipment was funded by BIS National E-infrastructure capital grant ST/K00042X/1, STFC capital grant ST/H008519/1, and STFC DiRAC Operations grant ST/K003267/1 and Durham University. DiRAC is part of the National E-Infrastructure. AB is supported by FCT-Portugal through grant SFRH/BD/75791/2011. MC thanks the support of the ERC Advanced Investigator grant COSMIWAY [grant number GA 267291]. The research leading to these results has received funding from the European Research Council under the European Union's Seventh Framework Programme (FP/2007-2013) / ERC Grant NuMass Agreement n. [617143]. This work has been partially supported by the European Union FP7  ITN INVISIBLES (Marie Curie Actions, PITN- GA-2011- 289442) and STFC. The bitbucket webpage of the {\tt COSMOSIS} code is https://bitbucket.org/joezuntz/cosmosis/wiki/Home.

\end{acknowledgments}

\appendix

\section{Parameter impact in the void fitting formula}

\begin{figure*}
	\centering
	\includegraphics[scale=0.40]{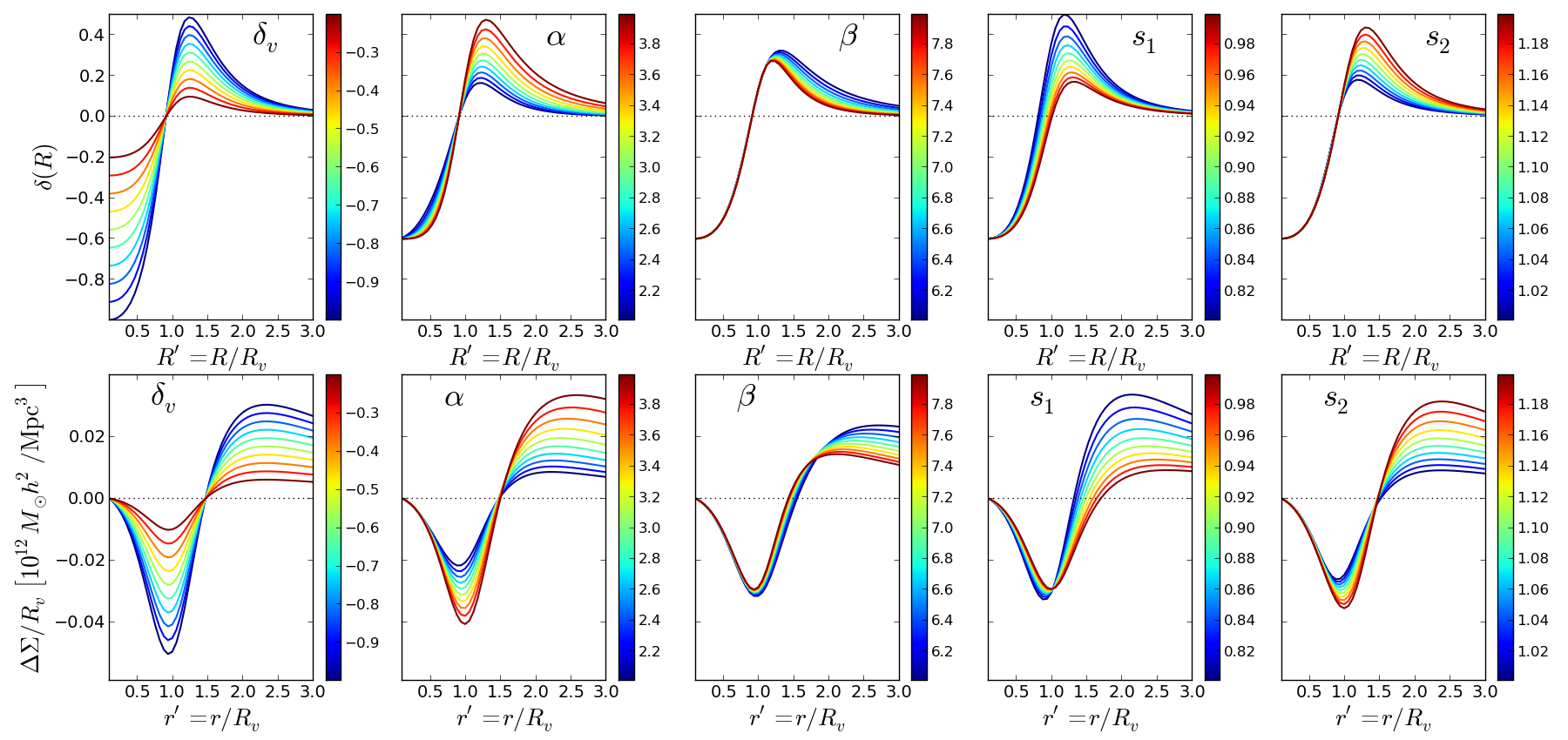}
	\caption{Impact of each of the parameters of the void density contrast formula of Eq.~(\ref{eq:mine}) (upper panels). The bottom panels show the respective lensing signal $\Delta\Sigma$. The curves are colored by the values of the parameter that is varying in each panel (from left to right, these are, $\delta_v$, $\alpha$, $\beta$, $s_1$ and $s_2$, respectively). The parameter values are indicated by the color bar at the right of each panel. When one parameter varies, the others are held fixed at their base values which are $\left( \delta_v, \alpha, \beta, s_1, s_2\right) = \left(-0.6, 3, 7, 0.9, 1.1\right)$. The calculation of the lensing signal was performed assuming GR.}
\label{fig:voidparamsimpact}\end{figure*}

Figure \ref{fig:voidparamsimpact} shows the effect that each of the five parameters $\left( \delta_v, \alpha, \beta, s_1, s_2\right)$ that enter Eq.~(\ref{eq:mine}) have on the void density and void lensing differential mass density profiles. The calculation of the lensing signal was performed with GR as the theory of gravity.

\bibliography{voidlensing.bib}

\end{document}